\documentclass[12pt,a4paper]{article}
\usepackage{graphicx}
\newcommand{\ii}{{\iota}}
\newcommand{\barn}{\bar{n}}
\newcommand{\n}{\hat{n}}
\newcommand{\msun}{\rm\,M_\odot}

\title{The Standard Cosmological Model and CMB Anisotropies}
\author{\small James G. Bartlett\\
	\small Observatoire de Strasbourg,
	\small 11 rue de l'Universit\'e,\\
	\small 67000 Strasbourg,  FRANCE\\ 
	\small {\tt bartlett@astro.u-strasbg.fr}\\
	\small Observatoire Midi--Pyr\'en\'ees, 14 ave. E. Belin\\
	\small 31400 Toulouse, FRANCE (after May 1, 1999)}

\begin{document}

\maketitle
\vspace{-1.5cm}
\begin{center}
\includegraphics*[height=5cm,width=5cm]{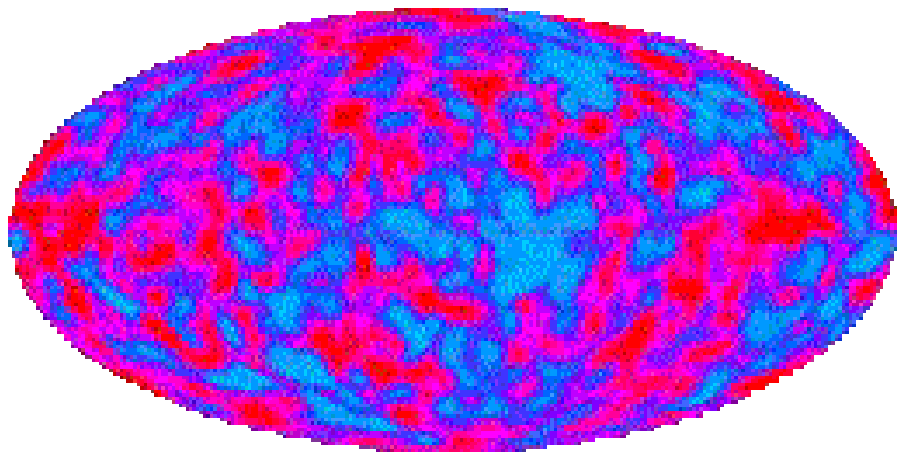}\\
\end{center}

This is a course 
on cosmic microwave background (CMB)
anisotropies in the standard cosmological model,
designed for beginning graduate students and advanced undergraduates.
``Standard cosmological model'' in this context means
a Universe dominated by some form of cold dark matter (CDM)
with adiabatic perturbations generated at some initial 
epoch, e.g., Inflation, and left to evolve
under gravity alone (which distinguishes it from
defect models).  The course is primarily theoretical 
and concerned with the physics of 
CMB anisotropies in this context and their relation 
to structure formation.  Brief presentations of
the uniform Big Bang model and of the observed 
large--scale structure of the Universe are given.
The bulk of the course then focuses on the 
evolution of small perturbations to the uniform
model and on the generation of temperature anisotropies
in the CMB.  The theoretical development is 
performed in the (pseudo--)Newtonian gauge because
it aids intuitive understanding by providing a quick
reference to classical (Newtonian) concepts.  The 
fundamental goal of the course is not to arrive at
a highly exact nor  exhaustive calculation of the 
anisotropies, but rather to a good understanding of the basic 
physics that goes into such calculations.

\newpage

\tableofcontents

\section{Introduction}

	The cosmic microwave background (CMB) is one of the cornerstones
of the homogeneous, isotropic Big Bang model.  {\em Anisotropies} in the
CMB are related to the small perturbations, superimposed on the perfectly
smooth background, which are believed to seed the 
formation of galaxies and
large--scale structure in the Universe.  The temperature anisotropies
therefore tell us about this process of large--scale structure formation.
For our purposes, the {\em standard cosmological model} shall be 
taken to mean the Big Bang scenario plus Inflation as the origin 
of perturbations.  Other models do exist; these are almost all 
exclusively built upon the Big Bang picture and only change 
the mechanism producing the perturbations (e.g., defect models,
presented by R. Durrer in this volume).

	In this set of lectures, we will be concerned with understanding
the production of CMB anisotropies in the standard model and 
what they can tell us about this model.  It is now well appreciated that
CMB anisotropies are exceptionally rich in detailed information of a
kind difficult to obtain otherwise.  Clearly, we expect them to
reflect the nature of the perturbations generated by Inflation and
which form galaxies by today.  Because perturbation evolution
depends on the constituents (dark matter, baryons, radiation, etc...),
we will also be able to learn something about these latter quantities and their
relative contributions to the total energy budget of the Universe.
In addition, the existence of a particle horizon, separating different
physical regimes, introduces a linear scale at last scattering
whose projected angular size gives us access to the global geometry of the
Universe.  With the wealth of data expected over the next decade,
from ground--based and balloon--borne instruments and, ultimately,
from the two planned satellites, MAP\footnote{\tt http://map.gsfc.nasa.gov/}
and Planck Surveyor\footnote{\tt http://astro.estec.esa.nl/Planck/}
(see Smoot 1997 for an excellent review of the CMB observation
programs, and \\
{\tt http://astro.u-strasbg.fr/Obs/COSMO/CMB/}
for a list of useful web links), one has 
honest reasons for expecting cosmology 
to take an important step forward in the next few years.

	The course has two essential and primary goals:
1/ to understand the physics responsible for 
the overall shape of the power spectrum
of density perturbations, $P(k)$ (Figure 1); and 2/
to understand the physics responsible for overall shape 
of the angular power spectrum of the CMB temperature 
anisotropies, $C_l$ (Figure 3).  In a nutshell,
we will explain these two figures
and the relation between them.  
We proceed by first briefly summarizing the salient
features of the Friedmann--Robertson--Walker model (Section
2) and of the observed large--scale structure (Section 3).  
We then move on into the bulk of the course and attack
our first goal by examining in detail the evolution of density 
perturbations in the expanding Universe (Section 4); this will
culminate in the construction of $P(k)$.  With this
physics in hand, we will finally be in a position to 
discuss the generation of CMB anisotropies, in Section 5,
which finishes with the construction of $C_l$.  
Our goal is not to exhaustively calculate to 
high accuracy CMB anisotropies, but to understand
the basic physics responsible for the density
and temperature anisotropy power spectra
predicted in the standard model.  This basic physics is a beautiful
application of General Relativity (GR), and the 
problems we seek to solve offer 
a magnificent stage upon which to observe the relationship 
between Newtonian and Einsteinian gravitational
dynamics.  Throughout, systematic use of the Poisson
gauge (Bertschinger 1996) will be used to highlight
this relationship.

	Finally, here is a list of some good general references:
\vspace{1cm}
\begin{itemize}
\item Bertschinger 1996, in {\em Cosmology and Large--scale
	Structure}, Les Houches session LX, Eds. R. Schaeffer et al. 
	(Elsevier:Amsterdam), p. 273
\item Bond 1996, in {\em Cosmology and Large--scale
	Structure}, Les Houches session LX, Eds. R. Schaeffer et al. 
	(Elsevier:Amsterdam), p.469
\item Efstathiou 1996, in {\em Cosmology and Large--scale
	Structure}, Les Houches session LX, Eds. R. Schaeffer et al. 
	(Elsevier:Amsterdam), p.133
\item Kolb E.W. and Turner M.S. 1990, The Early Universe, Frontiers in
	Physics, Addison--Wesley (Redwood City, CA)
\item Misner C.W., Thorne K.S. and Wheeler J.A. 1970, Gravitation,
	W.H. Freeman \& Co. (New York, NY)
\item Padmanabhan T. 1993, Structure Formation in the Universe,
	Cambridge University Press (Cambridge, UK)
\item Peebles P.J.E. 1993, Principles of Physical Cosmology, 
	Princeton Series in Physics, Princeton University Press (Princeton, NJ)
\item Peebles P.J.E. 1980, The Large--Scale Structure of the Universe,
	Princeton Series in Physics, Princeton University Press (Princeton, NJ)
\item Peebles P.J.E. 1970, Physical Cosmology,
	Princeton Series in Physics, Princeton University Press (Princeton, NJ)
\item Weinberg S. 1972, Gravitation and Cosmology, John Wiley \& Sons
	(New York, NY)
\end{itemize}

\section{Friedmann--Robertson--Walker Model}

	This section offers an extremely brief introduction to
the basics of Friedmann--Robertson--Walker (FRW) models; just
the essentials for what follows, such as the construction 
of comoving coordinates, the Friedmann equations and the
concept of the particle horizon.  We begin with the 
reasonable--sounding {\em cosmological principle}, first
coined by Einstein, that the Universe is, {\em in the  large}, 
{\bf homogeneous} and {\bf isotropic}.  Observationally, this
is supported by galaxy counts as a function of magnitude
(see, e.g., Peebles 1970, 1980), counts of galaxies in cells 
distributed throughout the Universe (from redshift surveys, e.g.,
Efstathiou 1996)
and, especially, the impressive isotropy of the CMB (apart from
the dipole, anisotropies only come in at $\sim 10^{-5}$). 
This comforts our aesthetic desire for the cosmological 
principle and permits us to deal with non--uniformity,
i.e., large--scale structure, as {\em small} perturbations to
an otherwise uniform FRW model.  This picture 
will be implicit in all that we do, and we will 
frequently refer to the unperturbed model as the
{\em background}, on which we follow the evolution
of small perturbations.

	Our concern in this section is the uniform FRW background.
The cosmological principle places strong mathematical constraints
on the permissible geometry of the spacetime of 
this background. Recall that in GR spacetime is a 
Riemann manifold with a 
metric tensor, ${\bf g}$, describing the fundamental, 
invariant line element, $ds$:
\begin{equation}
\label{eq:frw:metric}
ds^2 = g_{\alpha\beta}dx^\alpha dx^\beta
\end{equation}
The summation convention will be used throughout; 
Greek indices will be understood to run over (0,1,2,3), while
Latin indices only refer to spatial dimensions (1,2,3). 
The principle objects of the theory are the metric
coefficients, $g_{\alpha\beta}$.  Other important
quantities are the connection (Christoffel) 
coefficients
\begin{equation}\label{eq:frw:connect}
\Gamma^\alpha_{\beta \gamma} = \frac{1}{2} g^{\alpha \rho}
	\left( g_{\rho\beta, \gamma} + g_{\rho\gamma, \beta} 
	- g_{\beta\gamma, \rho} \right)
\end{equation}
the Ricci tensor
\begin{equation}
R_{\alpha \beta} = \Gamma^\rho_{\alpha \beta, \rho}
	- \Gamma^\rho_{\alpha \rho, \beta}
	+ \Gamma^\rho_{\gamma \rho} \Gamma^\gamma_{\alpha \beta}
	- \Gamma^\rho_{\beta \gamma} \Gamma^\gamma_{\alpha \rho}
\end{equation}
and the Ricci scalar, $R\equiv R^\alpha_\alpha$.
In these expressions, a comma (``{\bf ,}'') refers to a 
standard derivative with respect to the indicated variable.

	Because coordinates have no physical importance, physics
(via Einstein's field equations for gravity) only provides
six independent equations for the 10 elements of ${\bf g}$ 
(a symmetric tensor in four dimensions).  The remaining four
degrees of freedom must be {\em arbitrarily} imposed by a choice 
of coordinates.  This invariance of physics to general
coordinate transformations is known in GR as {\em gauge invariance}.
The standard form of the FRW metric is obtained 
by first imposing $g_{00}=1$ and
$g_{0i}=0$ (four conditions).  Thus, we need only be 
concerned with the the spatial metric, $^3g_{ij}$.  

	Notice that the cosmological principle is in fact
only a statement about {\em space}, i.e., that space--like
slices of 4D spacetime are homogeneous and isotropic;
it therefore only constrains
$^3g_{ij}$.  How so? Focus on the internal geometry of 
3--space by writing $^3g_{ij} = a^2(t) \tilde{g}_{ij}$,
where $\tilde{g}_{ij}$ is a time independent metric 
describing a general homogeneous and isotropic space 
(with Euclidean signature +++): 
$dl^2 = a^2(t)\tilde{g}_{ij}dx^idx^j$ 
Homogeneity tells us that this extraction of 
a time independent 3--metric is possible.
Isotropy implies that we may choose coordinates
such that
\begin{displaymath}
dl^2 = a^2(t)[f(r)dr^2 + r^2(d\theta^2+\sin^2\theta d\phi^2)]
\end{displaymath}
where the function $f(r)$ is yet to be determined.
This can be done by calculating the spatial curvature,
$\tilde{R} \equiv \tilde{R}^i_i$ -- the latter being the
spatial Ricci tensor obtained from the metric $\tilde{g}_{ij}$ --
and demanding that it be constant over all space (homogeneity).
By explicit calculation (hard work!), one finds 
that $f(r)=1/(1-\kappa r^2)$ guarantees a constant
curvature (of 3--space), with $\tilde{R}=6\kappa$.  
For a more abstract approach, see Weinberg (1972).

	We have found an expression
for the FRW metric coefficients and line element ($c=1$):
\begin{equation}\label{eq:frw:frwmetric}
ds^2 = dt^2 - a^2(t)\left[\frac{dr^2}{1-\kappa r^2} + r^2(d\theta^2
		+\sin^2\theta d\phi^2)\right]
\end{equation}
in a particular coordinate system known as {\em comoving 
coordinates}.
The function $a(t)$ is determined by the dynamics described in
the Einstein field equations.  The constant $\kappa$ is the
curvature of 3--space, which can be expressed in
terms of a comoving curvature radius, 
$R^2_{curv} \equiv 1/\kappa$. There are three 
important cases to distinguish:
\begin{enumerate}
\item $\kappa <0$, an infinite, hyperbolic space, leading to an {\em open}
	Universe
\item $\kappa =0$, an infinite, flat space, leading to a {\em critical}
	Universe
\item $\kappa >0$, a finite, spherical space, leading to a
	{\em closed} Universe ($r_{max} = R_{curv}$)
\end{enumerate}

	Consider now the motion of particles in this background.
Free--falling point masses follow {\em geodesic} paths
through spacetime
\begin{eqnarray}
\frac{du^\alpha}{ds} & + & \Gamma^\alpha_{\beta \gamma}u^\beta u^\gamma = 0\\
\nonumber
 & & u^\alpha \equiv \frac{dx^\alpha}{ds} \\
\nonumber 
& & u^2 = 1
\end{eqnarray}
where $u^\alpha$ is the four--velocity.  From the fact that
$\Gamma^i_{00} = 0$, we conclude that {\em comoving observers},
i.e., those who remain at constant $x^i$, are free--falling
observers following a specific geodesic.  Intuitively, this
had to be the case, for given that an observer at fixed $x^i$ 
sees homogeneity and isotropy around him/her, what direction
should he/she move in, if not fixed at constant $x^i$?!

	So far, this is all just kinematics of spacetime.  The dynamics
are described by Einstein's gravitational field equations ($c=1$):
\begin{equation}
G_{\alpha\beta} \equiv R_{\alpha\beta} -\frac{1}{2} g_{\alpha\beta} R 
	= 8\pi G T_{\alpha\beta}
\end{equation}
The tensor ${\bf G}$ is known as the Einstein tensor.
These are nonlinear, second order differential equations for 
the metric coefficients $g_{\alpha\beta}$, a normal occurrence
in classical dynamics.  They tell us that
geometry is ``sourced'' by stress--energy.
Notice that to leave four of the 10 metric coefficients
undetermined, which as mentioned corresponds to coordinate
freedom, there must be four constraints on the field equations.
These constraints are referred to as the Bianchi identities:
$(R^\alpha_\beta - 0.5 g^\alpha_\beta R)_{; \alpha}=0$, where the subscript 
``${\bf ;}$'' indicates a covariant derivative.   These same identities 
ensure energy conservation: $T^\alpha_{\beta ; \alpha}=0$.  
This highlights the beautiful connection, deeply ingrained in GR, 
between the coordinate invariance
of physics and energy conservation. 
 
	Our immediate concern is to find 
solutions for the scale--factor, $a(t)$, of the FRW metric.  To 
do so, we model the contents of the Universe with the stress--energy
tensor of an ideal fluid:
\begin{equation}\label{eq:frw:Tfluid}
T_{\alpha \beta} = (\rho + p)u_\alpha u_\beta - p g_{\alpha \beta}
\end{equation}
where $\rho$ and $p$ are the (constant) energy density and 
pressure of the fluid, and $u^\alpha$ is its four--velocity.
The assumed uniformity and isotropy (cosmological principle) demand the  
existence of a {\em universal} coordinate system in which 
$u^0=1$ and $u^i=0$, i.e., in which the fluid is {\em everywhere} at 
rest; this coordinate system will be the same as the comoving 
coordinate system introduced in the FRW metric, Eq. (\ref{eq:frw:frwmetric}).  
We can now
write down the field equations in terms of $a(t)$:
\begin{eqnarray}
\label{eq:frw:fried1}
\left(\frac{\dot{a}}{a}\right)^2 & = & \frac{8\pi G}{3}\rho - 
	\frac{\kappa}{a^2}\\
\label{eq:frw:fried2}
\frac{\ddot{a}}{a} & = & -\frac{4\pi G}{3} (\rho + 3p)
\end{eqnarray}
where an over-dot indicates a derivative with respect to cosmic time, $t$;
later in these lectures, notably when discussing 
relativistic perturbation theory, an over-dot will mean a derivative with
respect to {\em conformal time}, $\tau$ (Eq. \ref{eq:frw:tau}).
These are the time--time and space--space equations, the space--time
equations only leading to the useless identity $0=0$.  Eqs. 
(\ref{eq:frw:fried1}) and (\ref{eq:frw:fried2})
are known as the Friedmann equations.  The equation of 
energy conservation may be obtained either by combining 
the Friedmann equations, which according to the Bianchi 
identities must reproduce energy conservation, or directly
from $T^\alpha_{\beta ; \alpha}=0$:
\begin{equation}\label{eq:frw:energy_cons}
\dot{\rho} + 3\frac{\dot{a}}{a} (\rho + p) = 0
\end{equation}
The space component yields only $0=0$.  
It should be emphasized that only 2 equations
among (\ref{eq:frw:fried1}), (\ref{eq:frw:fried2}) and 
(\ref{eq:frw:energy_cons}) are 
independent.  This is insufficient to determine the
3 unknown functions $a(t)$, $\rho(t)$ and $p(t)$.
We need an additional equation -- the equation--of--state
for the matter, $p[\rho]$.  The system is closed by 
any {\em pair} of Eqs. (\ref{eq:frw:fried1}), (\ref{eq:frw:fried2}) 
and (\ref{eq:frw:energy_cons}), {\em and} the equation--of--state.  
This then leads to the familiar solutions: 
\begin{description}
\item[1] A Universe dominated by radiation, as appropriate
	at early times:
\begin{eqnarray*}
p & = & \frac{1}{3} \rho \\
\rho & \propto & 1/a^4  \\
a(t) & \propto &t^{1/2}
\end{eqnarray*}
\item[2] A Universe dominated by non--relativistic
	matter (dust), as appropriate after the radiation--matter
	transition:
\begin{eqnarray*}
p & = & 0  \\
\rho & \propto & 1/a^3  \\
a(t) & = & t^{2/3}
\end{eqnarray*}
\item[3] An open Universe dominated by curvature ($\kappa$--term), 
	perhaps the case today:
\begin{eqnarray*}
a(t) & \propto & t
\end{eqnarray*}
\item[4] A Universe dominated by a constant energy density,
	as appropriate during Inflation, and perhaps at present
	(cosmological constant):
\begin{eqnarray*}
a(t) & \propto & e^{Ht}
\end{eqnarray*}
where $H^2 = (8\pi G/3)\rho$ is a constant.
\end{description}

	A fundamental concept in these FRW models is that
of the particle horizon, or the distance that a photon has
been able to travel since the initial singularity.
Because photons travel along null geodesics, for which
$ds^2=0$, the corresponding comoving distance, $r_H$,
at a given cosmic time, $t$, is found from 
\begin{equation}\label{eq:frw:tau}
\int_0^{r_H} \frac{dr}{\sqrt{1-\kappa r^2}} = \int_0^t \frac{dt'}{a(t')} 
	\equiv \tau
\end{equation}
where $\tau$ is called the {\em conformal time}.
The {\em proper} horizon distance is $d_H = a(t)r_H$,
which for a critical Universe $= 2 t$ and $=3 t$ in the radiation-- and
matter--dominated eras, respectively (it is instructive
to make the same calculation for the Inflation epoch and
to reflect on the result).  The fact that the horizon
is proportional to the cosmic time seems, of course,
very reasonable.  Notice that for a flat Universe ($\kappa=0$), $\tau=r_H$.
At any given time, points separated 
by a distance larger than the horizon scale are not
in causal contact, and no causal physics (such
as pressure effects) can operate over scales
larger than the particle horizon.  This is an extremely
important aspect of the FRW background, and we
will see the central role that it plays in our studies.

\section{Large--scale Structure}

	We know of course that the Universe is not perfectly 
homogeneous and isotropic: there are galaxies and galaxy
clusters; the general large--scale distribution of 
galaxies is not random; and today we observe (finally!) 
temperature anisotropies in the CMB attesting to the
existence of deviations from perfect uniformity on
the largest scales.  These deviations
must represent real perturbations in the
density field of the Universe.  
Fortunately, apart from small scales, 
below $\sim 10 h^{-1}$ Mpc today (it is common in
describing large--scale structure to employ 
$h\equiv H_o/100$ km/s/Mpc), the perturbations are small
and may be treated by perturbation theory.  
In this section, we discuss how large--scale structure is
quantified.  This will be a short treatment of a vast
subject, designed only to lead us into the development of
the theory of small perturbations to the FRW background,
which we turn to in the next section.  The key 
concepts to take away from the discussion are  
that the density field is modeled as a 
{\em random field}, and that in the standard model it 
is completely characterized by the {\em power spectrum}.

	The fundamental question is how to quantify
the galaxy distribution, i.e., a distribution
of {\em points} in space. More specifically,
what do we mean by non--uniformity and how do we
give this concept a number?  Start with
an understanding of uniformity.
Notice that we are really interested
in tackling this question in a 
{\em statistical} sense -- it would seem 
absurd to suppose that uniformity should
correspond to a homogeneous lattice.
Rather, we imagine that the galaxies
were laid out with some randomness and
we want to understand if the responsible
mechanism operated in a uniform fashion.
Such a uniform process would have to give the same
{\em probability} of having a galaxy to
each position in space.  Dividing 
space into infinitesimal cells of volume $dV$, we expect
that the number of galaxies in each cell
is a Poisson random variable with
mean $=\barn dV$, where $\barn$ is a constant
number representing the mean density of galaxies.
Thus, {\em uniformity means a Poisson distribution
of points with constant mean density}.
 
 	By the term ``clustering'', we mean
the tendency of galaxies to group together
in space.  We can describe this as an
enhanced probability, over the uniform case,
of finding two galaxies in close proximity. 
This is how the two--point correlation function, $\xi_{gg}$,
is defined:
\begin{displaymath}
P_{12} = \barn^2 [1+\xi_{gg}(r)]dV_1 dV_2
\end{displaymath}
where $P_{12}$ is the probability of finding a galaxy in
a (infinitesimal) cell of volume $dV_1$ at position 1 {\bf and} another 
galaxy in a cell of volume $dV_2$ at position 2.  
Notice that if $\xi=0$, then $P_{12}$ is just
the product of the individual probabilities of
finding galaxies at points 1 and 2, as 
appropriate if the two events were independent;
a positive $\xi$ really does live up to its name
by measuring the statistical correlation between
the two random events of finding a galaxy at points
1 and 2. 

	An important remark: the correlation function
only depends on the {\em separation}
of the two positions (1 and 2), $r$, and not on their
location in space, nor on their relative 
orientation.  This is again the cosmological 
principle, in a new setting, showing up
as a requirement that the mechanism
producing galaxies operate in a {\em statistically}
homogeneous and isotropic fashion. 

	To make the connection to
density perturbations, let's first
rewrite the definition of the correlation
function in a suggestive form.
Consider again two cells of volume $dV_1$ and
$dV_2$ and their respective galaxy numbers, $N_1$ and
$N_2$.  If the cells are taken to be infinitesimally
small, so that $N_i=0$ or $1$, then these Poisson 
random variables satisfy 
$<N_1 N_2> = P_{12}$.  Introducing the 
local galaxy density contrast as $\delta_g
= [n(\vec{r}) - \barn]/\barn$, we may write 
\begin{displaymath}
<\delta_g(\vec{r_1}) \delta_g(\vec{r_2})> = \xi_{gg}(|r_2-r_1|)
\end{displaymath}
The fact that we expect $\delta_g$ to be 
somehow related to the {\em mass density contrast},
$\delta(\vec{r})\equiv [\rho(\vec{r})-\bar{\rho}]/\bar{\rho}$, 
then motivates us to consider this latter as a
{\em random field} with covariance
\begin{displaymath}
<\delta(\vec{r_1}) \delta(\vec{r_2})> = \xi(|r_2-r_1|)
\end{displaymath}
where this $\xi$ is the mass density 
two--point correlation function.  The
exact relation between the observed galaxy 
distribution and the actual, underlying 
mass density field is a fundamental question in 
large--scale structure studies, one which 
leads to the concept of {\em bias}.  A simple
linear bias is represented by $\xi_{gg} = b^2 \xi$,
where $b$ is called the bias factor.  This 
permits the galaxies to be more clustered than
the mass ($b$ is usually considered to be larger than 1).
Nature may demand a more complicated and non--linear
bias scheme; at present there very few restrictions on
the possibilities.  

	 We have been lead to the key idea 
that the mass density of the Universe is described 
by a {\em random field}, $\rho(\vec{r}) = \bar{\rho} + 
\delta\rho(\vec{r}) = \bar{\rho}[1+\delta(\vec{r})]$.  
This is the central 
concept underpinning all of large--scale structure
theory, where attention focuses on the density
contrast, $\delta(\vec{r})$.  It proves very 
useful to work with the modes of this field in 
Fourier space.  Our conventions, hereafter, will
be the following:
\begin{eqnarray*}
\delta(t,\vec{r}) & = & \frac{1}{(2\pi)^3} \int d^3k\; \delta_{\vec{k}}(t)
			e^{-i\vec{k}\cdot\vec{r}}\\
\\
\delta_{\vec{k}}(t) & = & \int d^3r\; \delta(t,\vec{r}) 
	e^{i\vec{k}\cdot\vec{r}}
\end{eqnarray*}
Wavenumbers, $k=2\pi/\lambda$, will expressed in terms of 
the {\bf comoving} wavelength.  
In Fourier space, the modes $\delta_{\vec{k}}(t)$ are
random variables with zero mean -- $<\delta_{\vec{k}}(t)>=0$ -- 
and covariance 
\begin{equation}\label{eq:lls:deltacov}
<\delta_{\vec{k}_1} \delta_{\vec{k}_2}> = (2\pi)^3 P(k) 
	\delta_D(\vec{k_1}-\vec{k_2})
\end{equation}
with 
\begin{equation}\label{eq:lss:Pk}
P(k) = \int d^3r\; \xi(r) e^{i\vec{k}\cdot\vec{r}}
\end{equation}
The function $P(k)$ is called the {\em power spectrum}, 
and we see that it is the Fourier Transform of the
two--point correlation function.  These relations
follow in a straightforward fashion from our 
Fourier conventions and the definition of $\xi$.
Notice that the fact that $\xi$ only depends on $r$ 
(the cosmological principle) implies that the power
spectrum is only a function of the magnitude $k=|\vec{k}|$.

	Quite often it is assumed that the density
perturbations are {\em Gaussian}, by which we mean
that the random variables $\delta_k$ [and $\delta(\vec{r})$]
are Gaussian random variables.  This is convenient, but
not necessary.  Inflation produced perturbations are 
expected to be Gaussian, so in the standard model these
are the only kind we consider.  Topological defect models,
on the other hand, can lead to non--Gaussian density
perturbations.  A set of Gaussian random variables
follow a multivariate Gaussian distribution, which 
is uniquely specified by a set of mean values and
a covariance matrix.  We have seen that, by 
definition, the mean values $<\delta_{\vec{k}}(t)>=0$, and further 
that the covariances are given by Eq. (\ref{eq:lls:deltacov}).
Our random variables are therefore independent (due to the
Dirac $\delta$--function) and completely specified by
the power spectrum, $P(k)$ (or by the two--point
correlation function, $\xi$).  For this reason,
in the standard model, {\em the power spectrum
is the fundamental theoretical quantity}. 
A set of cold dark matter (CDM) power spectra
is shown in Figure 1.  Our goal now is
to understand the origin of overall shape of
these power spectra, by studying perturbation
evolution, and then relate all to the CMB
anisotropies.

\section{Evolution of Density Perturbations}

	The structure seen in
the local Universe tells us that it is not perfectly 
homogeneous and isotropic, but that there are perturbations to
the totally uniform FRW model.  Fortunately, as we have
seen, these
may be considered small and treated by perturbation theory.
In the standard model, the perturbations
are created during Inflation, after
which they evolve only under the influence of gravity.
For this reason, they are often referred to as {\em passive}
perturbations.  This is in contrast to the {\em active}
perturbations generated at all times by topological
defects; in addition to gravity, these latter evolve under
the influence of non--gravitational forces, which
significantly complicates their treatment.  Restricting
ourselves to the standard model, we then need only be 
concerned with the gravitational evolution of 
perturbations.  This is an old subject in cosmology,
and it may be studied 
{\em independently of exact mechanism responsible for the
initial creation of the perturbations}; this is an
important simplification, which does not, for example,
apply to topological defect models. 

	In this chapter,
we first examine perturbation evolution in the Newtonian
approximation, where our physical intuition is more
easily satisfied (section 4.1).  Our results will apply
to sub-horizon perturbations ($a(t)\lambda<t$) of non--relativistic
fluids, such as the cold dark matter, and they
will enable us to understand some key physical processes 
shaping the power spectrum.   Since 
we are dealing with small perturbations, strong 
gravitational fields are not a limitation to the
Newtonian theory.  The limitation to the Newtonian
approximation instead comes from the existence of a 
particle horizon and from mass--energy equivalence
(e.g., the gravitational importance of pressure).  Thus, 
important situations 
that the Newtonian approximation is incapable of describing 
are the evolution of super-horizon 
perturbations ($a(t)\lambda>t$) and of perturbations in
relativistic fluids, such as the baryon--photon fluid 
around or before matter domination.  
Concerning super-horizon perturbations, observe that the
{\em proper} wavelength, $a(t)\lambda$, of a mode
grows like $a(t)\propto t^n$, where $n<1$ in the
radiation-- and matter--dominated eras.  On the 
other hand, the horizon grows proportionally to time, $t$,
and so as the Universe 
expands, the horizon gradually encompasses perturbations
of larger and larger wavelength.  Every mode was once larger 
than the horizon at early times.  We say such 
modes are ``outside'' the horizon and that they ``cross'' 
or ``enter'' the horizon when $\lambda_{prop} = a(t)\lambda
= t\sim H^{-1}$.  We turn
attention to the relativistic theory in Section 4.2.  
As mentioned, it is useful to refer to 
unperturbed spacetime (FRW) as the {\em background} and
imagine perturbations superimposed on and evolving in 
the background.  Peebles (1980) is an invaluable reference 
for all material found in this chapter;
Bertschinger (1996) is a particularly magnificent account 
of some of the relativistic theory.

\subsection{Non-relativistic approach}

{\noindent\bf Important: \em in this subsection, we 
denote comoving coordinates by $\vec{x}$ and proper
coordinates by $\vec{r}\equiv a(t)\vec{x}$.  Time is 
measured by the cosmic time $t$ (Eq. \ref{eq:frw:frwmetric}), and an 
over-dot indicates a derivative with respect to $t$.} \\

\noindent The Newtonian approximation may be applied when 
gravitational effects are weak, velocities and pressures 
are small, and changes to the gravitational field occur 
instantaneously.  This is generally the case in the
present--day Universe over regions with size, $R$, much 
smaller than both the horizon and the curvature radius
($R<< H^{-1}, R_{curv}$):  We see that the metric 
then becomes flat (Eq. \ref{eq:frw:frwmetric}),   
and if we further change from the comoving coordinates
used in the FRW metric, hereafter denoted by $\vec{x}$
in this subsection, to physical, proper coordinates,
defined by $\vec{r} = a(t)\vec{x}$, we find a truly
static, Newtonian background.  The restriction to regions
smaller than the horizon scale eliminates any effect due
to the finite propagation
time of changes to the gravitational field.  
Furthermore, we know that non--relativistic matter
dominates the energy density of the Universe after
the radiation--matter transition.
Thus, we attempt a classical fluid description:
\begin{eqnarray*}
\dot\rho(t,\vec{r}) + \nabla_r\cdot [\rho\vec{u}(t,\vec{r})] & = & 0\\
\\
\rho\nabla_r\Phi(t,\vec{r}) + \nabla_r p(t,\vec{r}) & = &
	-\rho\frac{d\vec{u}}{dt}\\
\ & = & -\rho[\dot{\vec{u}} + (\vec{u}\cdot\nabla_r)\vec{u}]\\
\\
\nabla_r^2\Phi & = & 4\pi G\rho\\
\end{eqnarray*}
where $\rho$, $p$ and $\vec{u}\equiv d\vec{r}/dt$ are the mass 
density, pressure
and (proper) velocity of the fluid, $\Phi$ is the gravitational 
potential, and all quantities are functions of both space and time.
The dot means time derivative at fixed $\vec{r}$, and all gradients,
taken with respect to the proper coordinates $r$,
are at fixed $t$.  

	Now comes a crucial point: we choose the
fundamental observers in this Newtonian description to 
correspond to those of the full FRW solution, i.e.,
in expansion with respect to each other and, therefore,
with fixed comoving coordinates $\vec{x}$.  These are
the galaxies, the observers who see a homogeneous and
isotropic Universe.  This may appear an obvious choice,
but that is only because we started with an understanding
of the full FRW solution to the equations of General 
Relativity; in fact, from a naive Newtonian perspective,
it is not at all obvious:  Why should {\em moving} observers
be the special ones who see homogeneity and
isotropy, instead of those at {\em rest in absolute space}
(to use a naive Newtonian language)?  
This choice escaped classical physicists, for whom the Universe had
to be unchanging, despite the fact that it would
have provided them with a coherent Newtonian model 
for the Universe.

	We wish to express our equations
in terms of the fundamental observers.  This means that
we should change from proper coordinates -- $(t,\vec{r})$ -- 
to comoving ones -- $(t,\vec{x})$ -- and work with
{\em peculiar velocity}, i.e., velocity with respect to the 
expansion: $\vec{v} = \dot{\vec{u}} - (\dot{a}/a)\vec{r}
= a\dot{\vec{x}}$.  Thus, all functions are to be expressed 
in terms of $(t,\vec{x}): f(t,\vec{x})=f(t,\vec{r}/a(t))$.  
By the chain rule
\begin{eqnarray*} 
\nabla_r|_t f(t,\vec{x}) & = & (1/a)\nabla_x|_t f(t,\vec{x}) \\
\\
\frac{\partial}{\partial t}|_r f(t,\vec{x}) & = &
	\frac{\partial}{\partial t}|_x f(t,\vec{x})
	-\frac{\dot a}{a} \vec{x}\cdot\nabla_x|_t f(t,\vec{x})\\
\end{eqnarray*}
With the understanding that $\nabla$ will hereafter 
mean a derivative with respect to $\vec{x}$
at fixed $t$, we rewrite the fluid equations in terms
of comoving coordinates, $(t,\vec{x})$, and peculiar velocity, 
$\vec{v}$, as
\begin{eqnarray}
\label{eq:ncosmo1}
\dot\rho + 3\frac{\dot{a}}{a}\rho + \frac{1}{a}\nabla\cdot(\rho\vec{v})
	& = & 0\\
\nonumber
\\
\label{eq:ncosmo2}
\ddot{a}\vec{x} + \dot{\vec{v}} + \frac{\dot{a}}{a}\vec{v}
	+\frac{1}{a}(\vec{v}\cdot\nabla)\vec{v} & = & 
	-\frac{1}{a}\nabla\Phi - \frac{1}{a\rho}\nabla p\\
\nonumber
\\
\label{eq:ncosmo3}
\nabla^2\Phi & = & 4\pi G\rho a^2 + [12\pi G p a^2]
\end{eqnarray}
These are the basic equations of Newtonian cosmology.
We have ``cheated'', in the purely Newtonian sense, by adding 
the term in square braces to the
last equation for the gravitational potential.  
This represents relativistic physics found only by taking
the Newtonian limit of the full Einstein 
field equations.  It is only in GR that we learn
that pressure can ``source'' gravity.

\subsubsection{Uniform solutions}   

	Before tackling perturbations, let's look at 
uniform solutions describing the background: 
$\rho = \rho_b(t)$, $p = p_b(t)$ and $\vec{v}=0$.  
Eqs. (\ref{eq:ncosmo1}),
(\ref{eq:ncosmo2}) and (\ref{eq:ncosmo3}) become
\begin{eqnarray*}
\dot{\rho_b}(t) + 3\frac{\dot{a}}{a}\rho_b & = & 0 \\
\\
\ddot{a}\vec{x} & = & -\frac{1}{a}\nabla\Phi_b \\
\\
\nabla^2\Phi_b & = & 4\pi G\rho_b(t) a^2 + [12\pi G p_b(t) a^2]\\
\end{eqnarray*}
Observe that the first equation is
the statement of matter conservation in an expanding
Universe: $\rho_b(t) \propto a^{-3}$.  The second equation
tells us that the fundamental observers experience an
acceleration caused by gravity.  Once again employing
a naive Newtonian vision, this acceleration is relative
to absolute space, or relative to the {\em one}
and {\em only} observer a rest in absolute space.
Placing this special observer at the origin and 
setting the potential to zero there, we see that
the equations describe the motion of 
matter shells centered on the origin (each
observer on the shell is equivalent, due to the
spherical symmetry assumed).  A less naive Newtonian 
picture is obtained by
incorporating the (weak) Galilean equivalence principle 
into a covariant description of Newtonian gravity.  
One then concludes that all fundamental 
observers are equal and each has the right to express these
equations relative to his/her own origin.  
A geometric theory of Newtonian gravity 
has been developed by Cartan 
(see, for example, Misner, Thorne and Wheeler 1970), and it
offers fascinating reading and a useful study for a deeper 
understanding of the nature of Einstein's theory 
of gravity.  

  	We find an equation
for the evolution of the scale factor, $a(t)$, by 
taking the divergence of the second equation and using the
third to eliminate the potential:
\begin{equation}
\label{eq:ncosmo:fried1}
\frac{\ddot{a}}{a} = -\frac{4\pi G}{3}\rho_b - [4\pi G p_b]
\end{equation}
This is one of the Friedmann equations (Eq. \ref{eq:frw:fried2}) .  
For zero pressure,
we can integrate it to obtain the other Friedmann equation
(\ref{eq:frw:fried1}):
\begin{equation}
\label{eq:ncosmo:fried2}
\left(\frac{\dot{a}}{a}\right)^2 = \frac{8\pi G}{3}\rho_b 
	 - \frac{\kappa}{a^2}
\end{equation}
where $\kappa$ is a constant of integration representing the
total (binding) energy of the expanding shell under consideration;
only in GR does it find its true calling as the 
curvature of space.  These equations follow from 
our choice of $p=0$ for the equation--of--state.
A situation in which pressure is important cannot
be completely treated from a Newtonian perspective, even with
the relativistic modification made to Eq. (\ref{eq:ncosmo3}), 
because the conservation law (\ref{eq:ncosmo1}), 
does not account for $p$ -- 
Newtonian physics does not recognize matter--energy 
conversion!  Later, we will find the relativistically
correct counterpart of Eq. (\ref{eq:ncosmo1}).  
To treat a pressure, or radiation, dominated background,
we must import, by ``hand'' into Eq.(\ref{eq:ncosmo:fried2}),
the relativistic result that 
$\rho_{rad}\propto 1/a^4$.

\subsubsection{Perturbations}

	Now turn attention to small perturbations away from 
perfect uniformity:  
\begin{eqnarray}
\label{eq:ncosmo:delta}
\rho(t,\vec{x}) & = & \rho_b(t) + \delta\rho(t,\vec{x})
			\equiv \rho_b(t) [1 + \delta(t,\vec{x})]\\
\nonumber
p(t,\vec{x}) & = & p_b(t) + \delta p(t,\vec{x})\\
\nonumber
\Phi(t,\vec{x}) & = & \Phi_b(t) + \phi(t,\vec{x})
\end{eqnarray}
The quantities $\delta$, $\delta p$, $\phi$ and 
the peculiar velocity, $v$,
are all taken to be small.  
Expanding the evolution equations to first order,  we obtain
\begin{eqnarray}
\label{eq:ncosmo:pert1}
\dot{\delta}(t,\vec{x}) + \frac{1}{a}\nabla\cdot\vec{v} & = & 0 \\
\nonumber
\\
\label{eq:ncosmo:pert2}
\dot{\vec{v}}(t,\vec{x}) + \frac{\dot{a}}{a}\vec{v} & = & 
	-\frac{1}{a}\nabla\phi - \frac{1}{a\rho_b}\nabla\delta p \\
\nonumber
\\
\label{eq:ncosmo:pert3}
\nabla^2\phi & = & 4\pi Ga^2\rho_b\delta
\end{eqnarray}
Assuming that the zero order (uniform) quantities are all given,
there are 4 unknown perturbation variables;
thus, we must add an equation--of--state ($p[\rho]$ or 
$\delta p[\delta\rho]$) to
close the system.  Various important physical regimes are
distinguished by different choices for the equation--of--state and
for the evolution of the background (itself related
to the zero order part of the equation--of--state). 

	By operating on the first equation with 
$(1/a^2)(\partial/\partial t) a^2$ and on the second with 
$a\nabla\cdot$, we may eliminate the velocity terms, producing
\begin{equation}
\label{eq:ncosmo:scalar}
\ddot{\delta}(t,\vec{x}) + 2\frac{\dot{a}}{a}\dot{\delta} = 
	4\pi G\rho_b(t)\delta + \frac{1}{a^2\rho_b}\nabla^2\delta p
\end{equation}
This is an equation of evolution for  {\em scalar perturbations}.
Why this name?  Because perturbations of this kind are 
described by scalar functions, even the velocity {\em vector}.
To see this, separate $\vec{v}$ into its parallel and transverse
components: $\vec{v} = \vec{v}_{\|} + \vec{v}_\bot$, with
$\nabla \times \vec{v}_\| = \nabla\cdot\vec{v}_\bot = 0$
(recall that this decomposition is always possible).  The
parallel component can be written in terms of a scalar function
as $\vec{v}_\| = \nabla f$, while the transverse component requires
another vector: $\vec{v}_\bot = \nabla\times \vec{A}$.  
Now, the equation we just
derived clearly eliminated $\vec{v}_\bot$ by the use of the
divergence to derive it.  For this reason, it is an equation
for a perturbation {\em mode} with only $\vec{v}_\|$ -- hence,
a scalar perturbation.  

	How about the vector mode, represented
by $\vec{v}_\bot$?  For this we must take the curl of the
second equation, finding
\begin{displaymath}
\frac{\partial}{\partial t} \nabla\times\vec{v}_\bot
	+ \frac{\dot{a}}{a}\nabla\times\vec{v}_\bot = 0
\end{displaymath}
The solution decays with the expansion factor as $1/a$, and
so these vector modes damp out.  Only scalar modes are 
{\em sourced} by (Newtonian) gravity.  This seems obvious in
hindsight -- after all, Newtonian gravity is a scalar field.
According to General Relativity, however, gravity is in
fact a tensor field, and so in the relativistic theory 
we may expect to find
tensor and vector modes, in addition to scalar perturbations.  
The tensor modes are gravity waves, 
a totally non--Newtonian phenomenon.  We shall see, however, that 
even in the relativistic context, only scalar perturbations
are sourced.  This is one of the fundamental 
differences between the standard model and topological 
defect scenarios, in which both vector and
tensor modes are actively sourced by the action of the
defects.   
This terminology of scalar, vector (and tensor) modes
is not usual in the Newtonian context, but it is helpful
to introduce it here, in familiar territory, before
encountering it in the full relativistic theory, where
it is commonly employed.

	Consider now specific solutions for
the (scalar) density perturbations described by Eq. 
(\ref{eq:ncosmo:scalar}).
Five are of particular interest, of which the first four
concern pressureless perturbations ($\delta p=0$) in different 
background models:
\begin{description}
\item[IA] $\Omega_{mat}=1$, i.e., matter--dominated background: 
	$\rho_b\propto 1/a^3$
\item[IB] $\Omega_{rad}=1$, i.e., radiation--dominated background: 
							 $\rho_b\propto 1/a^4$
\item[IC] $\Omega_{vac}=1$, i.e., cosmological constant domination: 
	$\rho_b=const$
\item[ID] $\Omega=0$, i.e., curvature--dominated background: $\rho_b=0$.
\end{description}
The fifth physical situation allows for non--zero matter
pressure
\begin{description}
\item[II] $p_{matter} \neq 0$
\end{description}
We may notice in advance a special property of solutions in 
the first four situations: the equation for $\delta$ contains
no spatial derivatives, and so the solutions will be
separable as $\delta(t,\vec{x}) =  A(\vec{x})\delta_g(t)
+ B(\vec{x})\delta_d(t)$, or in Fourier space as 
$\delta_k(t) =  A_k\delta_g(t)
+ B_k\delta_d(t)$, where $\delta_g$ and $\delta_d$
are the two independent solutions of Eq. (\ref{eq:ncosmo:scalar}).  
This means that in linear theory  
perturbations maintain their shape, changing only in amplitude 
(in these situations).  

	To obtain the solutions in each case, the procedure is always 
the same:  first find the background evolution -- $a(t)$ -- and 
then solve for the
perturbation.  Algebra is left to the reader; the results
are as follows:\\

\noindent{\bf IA: matter--dominated epoch}  
\begin{eqnarray*}
a(t) & \propto & t^{2/3}\\
\delta_g(t) & \propto & t^{2/3} \propto a(t) \\
\delta_d(t) & \propto & 1/t \propto 1/a^{3/2}(t) \\
\end{eqnarray*}
The important aspect of this solution, which may describe the
present--day Universe, is the existence of a growing mode
($\propto a(t)$).
\\

\noindent{\bf IB: radiation--dominated epoch}  
\begin{eqnarray*}
a(t) & \propto & t^{1/2}\\
\delta(t)_g & \propto & const \\ 
\delta_d(t) & \propto & -\ln(t)\\
\end{eqnarray*}
There is no growing mode in this case; in fact, a more
complete treatment incorporating both matter and radiation
and the matter--radiation transition finds result IA at late
times and a slow logarithmic growth during the radiation dominated
era (Peebles 1980).
\\

\noindent{\bf IC: vacuum--dominated epoch (Inflation)}  
\begin{eqnarray*}
a(t) & \propto & e^{Ht}, {\mbox{\rm \ \ \ with\ \ \ }} H=(\dot{a}/a)=const\\
\delta_g(t) & \propto & const \\
\delta_d(t) & \propto & 1/a^2\\
\end{eqnarray*}
Again, there is no growing mode.
\\

\noindent{\bf ID: curvature--dominated epoch ($\kappa <0$)}  
\begin{eqnarray*}
a(t) & \propto & t\\
\delta_g(t) & \propto & const \\
\delta_d(t) & \propto & 1/a \\
\end{eqnarray*}
Once again, no growing mode; Peebles (1980) develops the
solutions following through a matter--curvature transition.
\\

	Before moving on to the fifth and final case, let's consider when 
each of the above solutions might apply to the actual Universe.
The first case, {\bf IA}, applies to a critical
model, which accurately describes all scenarios after 
matter domination, but before either curvature or the cosmological
constant come to dominate (if ever).  Situation {\bf IB} 
represents the early Universe when radiation dominated
the total energy density, while cases {\bf IC} and {\bf ID} 
concern the late Universe when curvature or the cosmological
constant may become important.  Case {\bf IC} also 
describes the epoch of Inflation. 
In a Universe without a cosmological
constant, the curvature term begins to drive the expansion
at a redshift of approximately $(1+z_\kappa) \sim (1-\Omega)/\Omega$.
This can be found by comparing the two terms on the right--hand--side
of the Friedmann equation (\ref{eq:frw:fried1}).  Similarly, one sees 
that the cosmological constant dominates
after $(1+z_\Lambda)\sim (1-\Omega)^{1/3}/\Omega^{1/3}$ in a flat
Universe with $\kappa=0$.

	The last case, {\bf II}, is perhaps the most interesting,
because the spatial gradient remains in Eq. (\ref{eq:ncosmo:scalar})
and plays a crucial  
physical role.  As always, we must provide an equation--of--state.  
Before this was simply $p=0$; this time, we take a perfect
gas undergoing {\em adiabatic} (acoustic) oscillations:
\begin{equation}\label{eq:ncosmo:cs}
p = p_b + \delta p = p_b + \delta\rho\frac{\delta p}{\delta \rho}|_{adiabatic}
		 = p_b + c_s^2 \rho_b \delta 
\end{equation}
where the {\em sound speed}, $c_s$, is defined by $c_s^2 \equiv 
{\delta p}/{\delta \rho}|_{adiabatic}$. 
Assuming that the sound speed is constant, we find
\begin{displaymath}
\ddot{\delta}(t,\vec{x}) + 2\frac{\dot{a}}{a}\dot{\delta} = 
	4\pi G\rho_b(t)\delta + \frac{c_s^2}{a^2}\nabla^2\delta
\end{displaymath} 
The physical interpretation is simplest in Fourier space, where the
gradient operator is replaced by $-k^2$
\begin{displaymath}
\ddot{\delta_{\vec{k}}}(t) + 2\frac{\dot{a}}{a}\dot{\delta_{\vec{k}}} = 
	4\pi G\rho_b(t)\delta_{\vec{k}} - \frac{k^2c_s^2}{a^2}\delta_{\vec{k}}
\end{displaymath}
The competition between gravity and pressure 
is evident.  For long wavelengths (small $k$) the gravity term
dominates and causes the perturbation to grow.  Shorter wavelengths,
on the other hand, experience pressure resistance and tend
to oscillate.  The cross--over between the two regimes takes
place at $k_J $ such that $4\pi G \rho_b = c_s^2 k_J^2/a^2$.
Pressure has imposed a new scale, called the Jeans scale,
into the physics.  As a wavelength, this scale is expressed as
$a(t)\lambda_J = c_s(\pi/G\rho_b)^{1/2}$.
It is also commonly given as a mass: $M_J \equiv (4/3)\pi(a\lambda_J)^3
\rho_b$.

	We now have enough results to start painting a general 
picture of the evolution
of sub--horizon perturbations (for which this 
Newtonian approach works).  The essential point is that
{\em sub--horizon perturbations only grow in the matter--dominated phase};
unless $\Omega=1$, they are frozen at late times, and they are
always frozen at early times when radiation drives the 
expansion.  This has consequences for the shape of the density 
perturbation power spectrum.  As mentioned above, the horizon 
grows like the cosmic time $t$, while the proper wavelength of
a perturbation only grows like $a(t) \propto t^n$, where
$n<1$ (except during Inflation), and so that as the 
Universe expands, ever larger wavelength modes enter the
horizon. Thus, we see that as a perturbation enters the 
horizon during the radiation era, its amplitude is frozen 
to the value at horizon crossing.  Once in the matter era,
it starts to grow until (possibly) curvature and/or 
the cosmological constant begin control the expansion.  
	
\subsection{Relativistic approach}

{\bf Important: in this section we use the conformal time $\tau$,
and all time derivatives, represented by an over-dot, are 
taken with respect to $\tau$.}\\

	The full General Relativistic theory is needed
to accurately describe certain aspects of perturbation evolution.
Of particular importance to us is  the evolution of super-horizon
and pressure--dominated perturbations.  Our goal in this Section
is to develop an understanding of the workings of relativistic
perturbation theory, with the specific purpose of understanding
these two situations.
As mentioned, one of the main differences with 
Newtonian theory is the
tensor nature of gravity.  In principle, this leads to
scalar, vector and tensor perturbations.  All of these are
present in topological defect scenarios, but as we shall
see here, only scalar perturbations are sourced in the
absence of non--gravitational physics.  Even in the standard model,
however, gravitational waves produced by quantum fluctuations 
during Inflation, or perhaps at the Planck era, could be present 
and contribute to the anisotropies; 
on the other hand, vector perturbations only decay
with time.  Thus, in the standard model,
one in general considers both scalar and tensor perturbations.  

	We begin our study of relativistic perturbation
theory with the metric.  Once again, unperturbed 
spacetime will be referred to as the background, 
which is described by the FRW metric, Eq. (\ref{eq:frw:frwmetric}).
To keep things simple, we restrict ourselves to 
flat cosmologies, $\kappa=0$.  This is not
a severe restriction, because at the early times
of interest to us, all models are effectively flat.
In addition, it proves convenient to put space and time
coordinates on an equal footing in the background by
introducing the conformal time, $d\tau \equiv dt/a(t)$,
which, when $\kappa=0$, is the same thing as the comoving particle horizon,
Eq. (\ref{eq:frw:tau}).
The background line element is then 
\begin{equation}
\label{eq:cmetric}
ds^2 = a^2(\tau)[d\tau^2 - \gamma_{ij}dx^idx^j]
\end{equation}
where $\gamma_{ij} \equiv \tilde{g}_{ij}$ with $\kappa=0$.
In the following, we will often use $x$ to denote
the entire set of comoving spacetime coordinates, i.e., $x\leftrightarrow 
(\tau,\vec{x})$.

	We may write the most general {\em perturbed} metric as
\begin{eqnarray}
\label{eq:pmetric}
ds^2 & = & a^2(\tau)\left[(1+2\psi)d\tau^2 - \omega_id\tau dx^i
	- [(1-2\phi)\gamma_{ij} + 2h_{ij}]dx^idx^j\right] \\
\nonumber
\\
\nonumber
\gamma_{ij}h^{ij} & = & \gamma^{ij}h_{ij} = 0 \quad\quad{\rm (traceless)}
\end{eqnarray}
In this expression, $\psi(x)$ and $\phi(x)$ are scalar
fields, $\omega_i(x)$ is a vector field and $h_{ij}(x)$ is a 
tensor field, {\em all defined on the background {\bf 3--space}}.
These quantities are functions of $\tau$ and $\vec{x}$, but
their transformation properties are defined on the 3--space
manifold; notice, for example, that $\omega_i$ is a 3--vector
(not a 4--vector).  For this reason, indices on
$\omega_i$ and $h_{ij}$ will be raised and lowered using the
3--metric $\gamma_{ij}$ (in technical terms, the isomorphism
between forms and vectors is given by the 3--metric).  Use of
only the unperturbed part of the 3--metric is justified when
working to first order.
That this is the most general perturbed metric can
be seen from the fact that there are ten separate 
degrees--of--freedom, as required for a general metric:
$\psi,\phi$ (2), $\omega_i$ (3) and $h_{ij}$ (5).
The tensor $h_{ij}$ has only five independent elements,
because its trace has been explicitly put in $\phi$.
These scalar, vector and tensor quantities are not
yet the true scalar, vector, tensor perturbations.  Remember that
a vector, such as $\omega_i$, has both scalar and vector parts.
The tensor, $h_{ij}$, may be broken down to its
scalar, vector and tensor parts as follows:
\begin{eqnarray*}
{\bf h} & = & {\bf h}_{||} + {\bf h}_\bot + {\bf h}_T \\
\\
h_{ij\ ||} & = & (\partial_i\partial_j - \frac{1}{3}\gamma_{ij}\nabla^2) h\\
\\
h_{ij\ \bot} & = & h_{i,j} + h_{j,i}\\
\\
h_{ij\ T}^{\ \ ,j} & = & 0
\end{eqnarray*} 
where $h$, $h_i$ and ${\bf h}_T$ are scalar, vector ($\partial^i h_i=0$)
and tensor functions, respectively.  These are the functions we 
mean when speaking about scalar, vector and tensor
perturbations.

	The perturbed metric has, at this point, ten
components, but as discussed earlier, there are only
six independent Einstein field equations.  We must
fix the remaining four degrees--of--freedom by a 
choice of coordinates.  Recall that the invariance of
physics to general coordinate transformations is formally
known in GR as gauge invariance.
For this reason, a choice of coordinates is also
called a {\em gauge}.  One is not obliged to choose a
particular gauge to develop perturbation theory; there
are explicitly covariant approaches (see, e.g., Bertschinger
1996 and references therein).  If, on the
other hand, one decides to fix a gauge, then there are
an infinity of choices, of which two are commonly employed:
\begin{eqnarray}
\nonumber
\label{eq:sgauge}
\psi \equiv \omega_{i} \equiv 0 & & {\rm Synchronous\ Gauge}\\
\nonumber
\\
\label{eq:pgauge}
\omega_i^{\ ,i} \equiv h_{ij}^{\ \ ,j} \equiv 0 & & {\rm Poisson\ Gauge}
\end{eqnarray} 
The first is used, for example, by Peebles (1980), while the
second is developed at length by Bertschinger (1996).
For the time being, we will proceed in 
full generality, specifying a gauge only when necessary.
Our immediate goal is to write down the conservation
equations ($T^{\alpha\beta}_{\ \ \ ;\beta}=0$) and the field
equations ($G_{\alpha\beta} = 8\pi G T_{\alpha\beta}$) in terms
of the perturbation variables.  This is a lot of work(!) and is 
left to the reader as an exercise well worth the effort.
Here, we outline the major steps and give the important 
intermediate results required to obtain the final
expressions.

	Firstly, we write down the inverse metric coefficients
to first order:
\begin{eqnarray*}
g^{00} & = & \frac{1}{a^2}(1-2\psi) \\
g^{0i} & = & -\frac{1}{a^2}\omega^i \\
g^{ij} & = & -\frac{1}{a^2}[(1+2\phi)\gamma^{ij} - 2h^{ij}]
\end{eqnarray*}
With these in hand, and after much algebra, we 
find the connection coefficients, Eq. (\ref{eq:frw:connect}):
\begin{eqnarray}\label{eq:rcosmo:gamb}
\Gamma^0_{00} & = & \frac{\dot{a}}{a} + \dot{\psi} \\
\Gamma^0_{0i} & = & \psi_{,i} + \frac{\dot{a}}{a}\omega_i \\
\Gamma^0_{ij} & = & -\frac{1}{2}(\omega_{i,j} + \omega_{j,i})
	+ \frac{\dot{a}}{a}[(1-2\phi)\gamma_{ij} + 2h_{ij}]
	+ \dot{h}_{ij} - (\dot{\phi}+2\frac{\dot{a}}{a}\psi)\gamma_{ij} \\
\Gamma^i_{00} & = & \frac{\dot{a}}{a}\omega^i + \dot{\omega}^i 
	+ \psi^{,i} \\
\Gamma^i_{0j} & = & \frac{1}{2}(\omega^i_{\ ,j} - \omega_j^{\ ,i})
	- \dot{\phi}\delta^i_j + \dot{h}^i_j 
	+ \frac{\dot{a}}{a}\delta^i_j\\
\label{eq:rcosmo:game}
\Gamma^i_{jk} & = & (h_{j, k}^i + h_{k ,j}^i - h_{jk}^{\ \ ,i})
	- \frac{\dot{a}}{a}\omega^i\gamma_{jk} 
	- (\phi_{,k}\delta^i_j + \phi_{,j}\delta^i_k - 
		\phi^{,i}\gamma_{jk})
\end{eqnarray}
Remember that the over-dot in these equations refers to a derivative
with respect to {\bf conformal time $\tau$}.

	As in Eq. (\ref{eq:frw:Tfluid}), we describe the cosmic 
plasma as a perfect fluid with stress--energy tensor
\begin{displaymath}
T_{\alpha\beta} = (\rho+p)u_\alpha u_\beta - pg_{\alpha\beta}
\end{displaymath}
where $u^\alpha \equiv dx^\alpha/ds$ is the fluid's 
4--velocity ($ds$ is the proper time 
in the fluid's rest frame at the point in question).  
Thus, $u^0 = d\tau/ds$ and $u^i=u^0 dx^i/d\tau \equiv u^0v^i$.
The {\em peculiar velocity} of the fluid, $v^i$, is a first
order perturbation variable. From the condition 
$u^\alpha u_\alpha = 1$, we find $u^0 = (1-\psi)/a$ to
first order; thus, $u^i = v^i/a$ to first order.  The 
covariant components are $u_\alpha = g_{\alpha\beta}u^\beta$:
$u_0 = (1+\psi)a$ and $u_i = -a(\omega_i+v_i)$, where
$v_i\equiv \gamma_{ij}v^j$. 
This permits us to write the following 
expressions for the stress--energy tensor
to first order in the perturbation variables:
\begin{eqnarray*}
\label{eq:sepert1}
T^0_0 & = & \rho = \rho_b + \delta\rho \\
\nonumber\\
\label{eq:sepert2}
T^i_0 & = & (\rho_b+p_b)v^i \\
\nonumber\\
T^i_j & = & -p\delta^i_j = -(p_b+\delta p)\delta^i_j
\end{eqnarray*}

	We are now in a position to write down the components of 
the conservation law ($T^\beta_{\alpha; \beta}=0$):
\begin{eqnarray}
\label{eq:claw0}
\dot{\rho} + 3\left(\frac{\dot{a}}{a}-\dot{\phi}\right)(\rho+p)
	+ \partial_i[(\rho+p)v^i] & = & 0 \\
\label{eq:clawi}
\partial_\tau[(\rho+p)(\omega_i+v_i)] + 4\frac{\dot{a}}{a}
	(\rho+p)(\omega_i+v_i) + \partial_ip + (\rho+p)\psi_{,i} & = & 0
\end{eqnarray}
It must be remembered that in these formulae $\rho=\rho_b+\delta\rho$
and $p=p_b+\delta p$, and that the expressions are only valid
to {\bf first order}.
These equations are the relativistic
generalizations of the fundamental Newtonian relations used in
the last section, Eqs. (\ref{eq:ncosmo1}) and 
(\ref{eq:ncosmo2}).  
It is instructive compare the unperturbed version
of these equations, which describe the background,
to their Newtonian counterparts.  For a 
uniform fluid, Eq. (\ref{eq:claw0})
reduces to
\begin{equation}
\dot{\rho}_b + 3\frac{\dot{a}}{a}(\rho_b+p_b) = 0
\end{equation}
For zero pressure, we find recover the Newtonian continuity
equation, as expected.  If $p=\rho/3$, as appropriate for radiation,
we find the result that $\rho_{rad}\propto 1/a^4$, which
we had to import ``by hand'' into the Newtonian theory to treat 
a radiation--dominated plasma.  Interestingly, the space part reduces 
to the identity $0=0$.  
The first order components of the conservation law 
for {\bf adiabatic perturbations} are
\begin{eqnarray}
\label{eq:clawpertt}
\dot{\delta} + 3(c_s^2-x)\frac{\dot{a}}{a}\delta + (1+x)(\partial_i v^i
	-3\dot{\phi}) & = & 0\\
\label{eq:clawperts}
\frac{1}{a^4}\partial_\tau
	\left[a^4\rho_b(1+x)(\omega_i+v_i)\right]
	+ c_s^2\rho_b\partial_i\delta + \rho_b(1+x)\psi_{,i} & = & 0
\end{eqnarray}
In these equations, $\delta$ refers to the energy density
contrast (Eq. \ref{eq:ncosmo:delta}), $x\equiv p_b/\rho_b$ and, as 
before, the sound speed is $c_s$ (Eq. \ref{eq:ncosmo:cs}).
These two relations are the relativistic generalizations of
Equations (\ref{eq:ncosmo:pert1}) and (\ref{eq:ncosmo:pert2}),
as will become clear shortly.

	To study the behavior of the fluid, 
we also need the Einstein field equations for 
the perturbation variables; in other words, the
equivalent of the Poisson equation, (\ref{eq:ncosmo:pert3}).
Because GR is a tensor theory of gravity, there are many more
equations describing the dynamics than just a Poisson--like
relation.  The zero order field equations
were already found as the Friedmann equations; now,
we need the first order relations.  Hereafter, we restrict ourselves to the
{\bf Poisson gauge}.  Remembering the gauge conditions, Eq (\ref{eq:pgauge}),
we find the following first order equations for 
{\bf adiabatic perturbations}:
\begin{equation}\label{eq:rcosmo:pert1}
\nabla^2\psi  +  3\left[\ddot{\phi} +\frac{\dot{a}}{a}
	(\dot{\phi}+\dot{\psi})\right] - 6\left[\frac{\dot{a}^2}{a^2}
	-\frac{\ddot{a}}{a}\right]\psi =   4\pi Ga^2\rho_b
	(1+3c_s^2)\delta 
\end{equation}
\begin{equation}\label{eq:rcosmo:pert2}
\nabla^2\omega_i  =  16\pi G a^2 \rho_b (1+x)(\omega_i+v_i)_{\bot}
\end{equation}
\begin{equation}\label{eq:rcosmo:pert3}
\dot{\phi}_{,i} + \frac{\dot{a}}{a}\psi_{,i} = -4\pi Ga^2\rho_b
	(1+x)[v_i]_{||}
\end{equation}
\begin{equation}\label{eq:rcosmo:pert4}
\ddot{\phi}  -  \nabla^2\phi-\frac{1}{3}\nabla^2(\phi-\psi)
	+\frac{\dot{a}}{a}(5\dot{\phi}+\dot{\psi})
	+ 2\left[\frac{\ddot{a}}{a}+\frac{\dot{a}^2}{a^2}\right]\psi
	= -4\pi Ga^2\rho_b(1-c_s^2)\delta
\end{equation}
\begin{equation}\label{eq:rcosmo:pert5}
(\phi-\psi)_{,ij}  -  \frac{1}{3}\gamma_{ij}\nabla^2(\phi-\psi)  =  0 
		\hspace*{1cm} (i\ne j)
\end{equation}
\begin{equation}\label{eq:rcosmo:pert6}
\left(\partial_\tau + 2\frac{\dot{a}}{a}\right)
	(\omega_{i,j}+\omega_{j,i})  =  0 
\end{equation}
\begin{equation}\label{eq:rcosmo:pert7}
\left(\partial^2_\tau - \nabla^2 + 2\frac{\dot{a}}{a}\partial_\tau\right) 
	h_{ij}  = 0
\end{equation}
They have been separated into their scalar, vector and 
tensor parts, each of which evolves independently of the others;
this is the value of the decomposition. 
From the last three equations, we learn that $h_{ij}$ represents
gravity waves evolving according to a homogeneous equation with
damping due to the expansion; that the vector mode, $\omega_i$,
is damped by the expansion, as promised; and that $\psi=\phi$.
If $\vec{\omega}$ is damped, then the second equation tells
us that so is $\vec{v}_{\bot}$, again as expected from 
our Newtonian exploits.  
Employing these results and combining the first and fourth
equations, we find a nice expression for the ``Newtonian potential'',
$\psi$:
\begin{equation}\label{eq:rcosmo:apoisson}
\nabla^2\psi - 3\frac{\dot{a}}{a}(\dot{\psi}+\frac{\dot{a}}{a}\psi) 
	= 4\pi Ga^2\rho_b\delta
\end{equation}
which looks encouragingly like the Poisson equation for 
a Newtonian gravitational field (hence its name
in the cosmological setting).  

	Let's use these equations to once again study the evolution
of density perturbations.  Since these are scalar perturbations,
we shall hereafter ignore $\vec{\omega}$ and $h_{ij}$.  Notice
that we then only have three independent equations among 
(\ref{eq:clawpertt}), (\ref{eq:clawperts}),
(\ref{eq:rcosmo:pert1}), (\ref{eq:rcosmo:pert3}) and
(\ref{eq:rcosmo:pert4}), which we will use in form
of Eqs. (\ref{eq:clawpertt}), (\ref{eq:clawperts}) and
(\ref{eq:rcosmo:apoisson}).  The requisite equation--of--state
is embodied in the variable $x$.  Remember, our ultimate goal is
to reconstruct the general features of the power spectrum.
We have already gotten quite far in this direction with
our Newtonian analysis, which permited us to study density
perturbations of a non--relativistic fluid (such as the
CDM) inside the horizon.  Two important 
situations not covered by the Newtonian approach
are the evolution of a radiation--dominated fluid,
such as the baryon--photon fluid,
and the evolution of perturbations outside the horizon,
before they (re)enter.  It is here that the relativistic
equations are needed.  

	Consider first the question
of super-horizon modes.  In the limit $\lambda >\tau$, spatial
gradients may be ignored relative to time derivatives and
terms with $\dot{a}/a$.  This reduces the conservation
equations (\ref{eq:clawpertt}) and
(\ref{eq:clawperts}) and the ``Poisson'' equation 
(\ref{eq:rcosmo:apoisson}) to 
\begin{eqnarray}
\label{eq:rcosmo:lspert2}
\dot{\delta} + 3(c_s^2-x)\frac{\dot{a}}{a}\delta -  
	3(1+x)\dot{\psi} & = & 0\\
\partial_\tau\left[a^4\rho_b(1+x)\vec{v}\right] & = & 0\\
\label{eq:rcosmo:lspert1}
\dot{\psi} + \frac{\dot{a}}{a} \psi & = & -\frac{1}{2}
	\frac{\dot{a}}{a}\delta
\end{eqnarray}  
We have used the Friedmann equations in terms of conformal time
to obtain the last expression.
The second equation tells us that the peculiar velocity is damped
as $1/a$ in the matter--dominated epoch, but constant
during the radiation--dominated phase.
By taking the time derivative of the last equation, and
then using the first relation to eliminate $\delta$ and 
the Friedmann equations to eliminate $\ddot{a}$, one
obtains the evolution of $\psi$:
\begin{displaymath}
\ddot{\psi} + 3(1+c_s^2)\frac{\dot{a}}{a}\dot{\psi} 
	+ 3(c_s^2-x)\left(\frac{\dot{a}}{a}\right)^2\psi   = 0
\end{displaymath}
There are no growing mode solutions to this
equation -- at best a constant mode when $x=c_s^2$,
applicable to the cosmic fluid
during the radiation-- and matter--dominated epochs,
in addition to a decaying
solution; this also implies that 
$\delta$ is constant during these two epochs.  
The special case of $x=-1$,
applicable during the epoch of Inflation, leads as well to 
a constant solution, because this is the only way to 
satisfy Eqs (\ref{eq:rcosmo:lspert1}) and
(\ref{eq:rcosmo:lspert2}).  On the other hand, the potential decays
in the transitions between these periods.  In summary, we have that
{\bf the perturbation variables $\delta$ and $\psi$ remain
constant outside the horizon during the vacuum--, radiation--
and matter--dominated epochs}.  This is a useful result,
but remember that it is a {\em gauge dependent} statement, valid
only in the Poisson gauge.  It is not a general statement because these
variables are not gauge independent quantities.  

	Now turn attention to the small--scale
limit of the GR perturbation equations ($\lambda < \tau$).  This 
will give us the evolution of a pressure--dominated fluid,
the second relativistic case we need to fully understand
the power spectrum.  Recalling that $a^2\rho_b$ 
is $\sim 1/\tau^2$ (Friedmann equations,
in terms of conformal time), Eq. (\ref{eq:rcosmo:pert3})
tells us that $\dot{\psi}+(\dot{a}/{a})\psi$ is of order $\lambda/\tau^2$, 
and hence the second term in Eq. (\ref{eq:rcosmo:apoisson})
may be dropped relative to the Laplacian -- we recover Poisson's
equation for the Newtonian potential.  Next, the conservation
equations:  We have just argued that $\dot{\psi}\sim \lambda/\tau^2$,
so the third term in Eq. (\ref{eq:clawpertt})
is dominated by the divergence, leaving 
\begin{displaymath}
\dot{\delta} + (1+x)\nabla\cdot\vec{v} + 3(c_s^2-x)\frac{\dot{a}}{a}
	\delta = 0
\end{displaymath}
In the limit of zero pressure, $x=c_s^2=0$, we recover the classical
mass conservation law.
Don't forget that here our over-dots mean {\bf conformal time}
derivatives, which explains why the form is not exactly as
presented in Eq. (\ref{eq:ncosmo:pert1}).  
The relativistic generalization of 
the Euler equation (\ref{eq:ncosmo:pert2}) 
is obtained from Eq. (\ref{eq:clawperts}) using
similar reasoning; one finds
\begin{displaymath}
\dot{\vec{v}} + (1-3x)\frac{\dot{a}}{a}\vec{v} = -\frac{c_s^2}{1+x}
	\nabla\delta - \nabla\psi
\end{displaymath}
Once again, the classical result is obtained for $x=c_s^2=0$.
These equations generalize the Newtonian results by 
properly taking into account the gravitational effects
of pressure.  From them, we see that even in the radiation--dominated
epoch ($x=c_s^2=1/3$), perturbations in the (relativistic) baryon--photon
fluid oscillate as sound waves, just as they did in our more
classical treatment (which, by the way, never applies to the
this fluid which remains essentially radiation dominated up til 
decoupling).  Thus, we have treated
the two particular cases of interest which were not
properly treated by the Newtonian theory, i.e.,
super-horizon perturbations and a pressure--dominated fluid.

\begin{figure}\label{Pk}
\begin{center}
\includegraphics*[height=14cm,width=14cm]{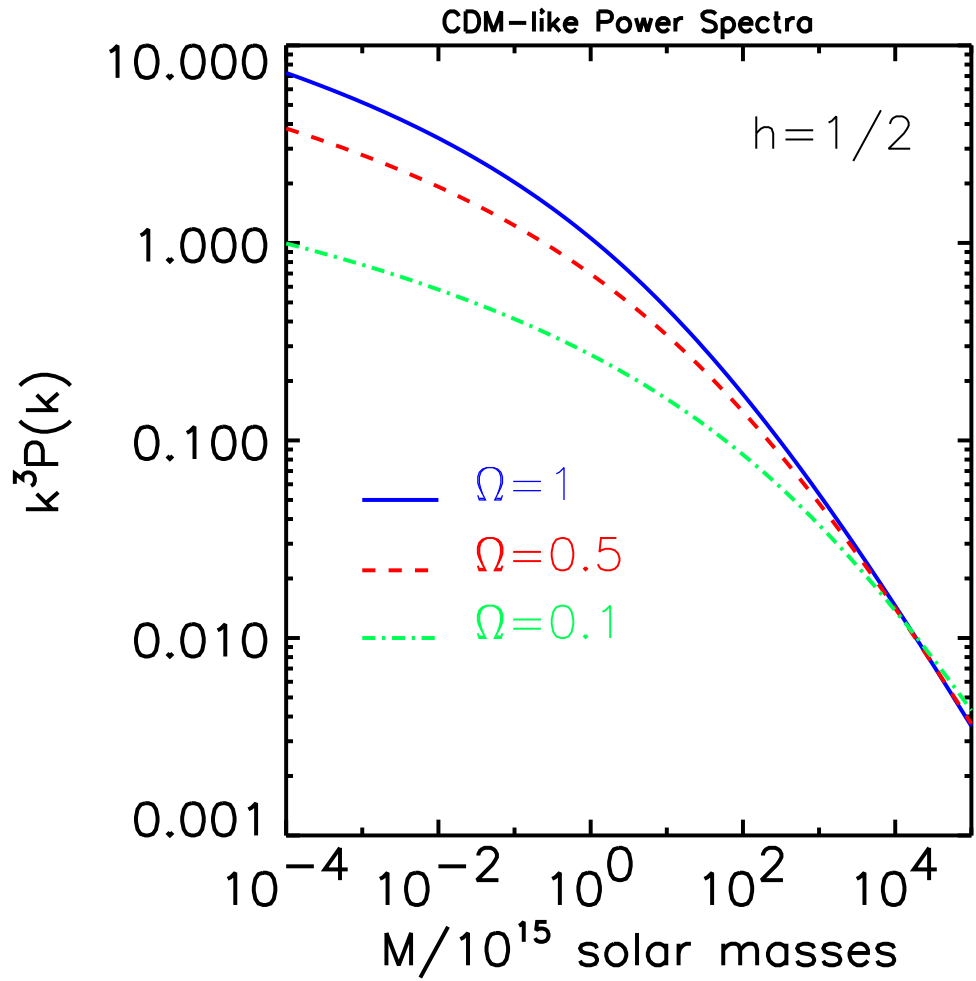}
\end{center}
\caption{Cold dark matter density power spectra.
The quantity $k^3P(k)$, approximately equal to the real--space 
density perturbation, is shown as a function
of mass scale, $M=(4\pi/3)(2\pi/k)^3\rho_b$.  
All spectra have been arbitrarily normalized on 
large scales.  The 
$x$--axis is given in units of $10^{15} \msun$.
The mass scale corresponding to the horizon at 
the radiation--matter transition is $\sim 10^{15}
(\Omega h^2)^{-2} \msun$, or $ 1.0(\Omega h^2)^{-2}$ in
these units.  From a power--law, all spectra gradually
break over to a constant on small scales; the roll--over
is very gradual.  The spectra for the three different
values of $\Omega$ demonstrate the effect of lowering
the cosmic mean density, and thereby increasing the
mass scale of the turnover.}
\end{figure}
 
\subsection{Constructing the density power spectrum}

	To finally understand the general form of CDM--like
power spectra using our previous results, we need 
to specify the primordial spectrum,
say from Inflation.  The simplest Inflation models
are based on a quantum field in which no important
physical scales are introduced (the fluctuating part
of the quantum field is effectively free)  -- the only scale in 
the problem remains the proper (event) horizon $H^{-1}$ (a constant).  
During inflation, the proper wavelength of a quantum
mode of the field grows exponentially
with the expansion to eventually become much bigger
than the horizon.  When $a\lambda = H^{-1}$,  
we say the perturbation ``crosses'' the
horizon, going out.  It then becomes a classical
perturbation mode subject only to gravitational
influences.  From what we have just said,
we should not be surprised to learn that every perturbation looks the same
as it crosses the horizon, because this is the only
scale in the physics.  In fact, the quantitative
result we are looking for is that the density or
gravitational potential perturbations crossing the horizon
are fixed in real space; in terms of the power spectrum,
this implies that $k^3P(k) = const$ at horizon crossing. 
Without too much extra complication, one can
imagine a more general spectrum given by a featureless
power--law: $k^3P(k)\propto k^\gamma$ at horizon
crossing.  This occurs, for example, in Inflation
models where the energy density of the vacuum slowly
decreases with time, resulting in a $\gamma <0$  
(this does imply the introduction
of an additional physical scale, namely the time constant for the
vacuum evolution).  

	After Inflation, the horizon grows faster than the
proper wavelength of a perturbation ($H^{-1}\sim t$ verses 
$a\lambda \sim a \sim t^n$, 
with $n<1$) and, one by one, the perturbations re-cross the
horizon going in.  Now, we have seen that in Poisson gauge 
super-horizon perturbations remain constant in the matter-- and
radiation--dominated epochs, i.e., $\dot{\psi}=\dot{\delta}=0$.  
If we ignore for simplicity their slight decay in the transition,
we obtain that each perturbation re-enters with the same
$k^3P(k)$ that it had going out.  Perturbations therefore
re--enter the horizon, at $\tau=\lambda$, with
either increasing or decreasing amplitude, depending
on the sign of $\gamma$.  This implies that there
must be a cut--off somewhere in the spectrum, because
otherwise we would be faced with a hard divergence on
either large or small scales.  The special case
of $\gamma=0$ avoids this (it has, in fact, a logarithmic
divergence, which is much easier to treat) and
leads to what is called a {\em scale invariant} spectrum, 
since each scale re--enters the horizon with the same amplitude.  
Because of its aesthetic properties, this scale invariant spectrum was 
postulated well before the concept of Inflation and is
often referred to as the Harrison--Zel'dovich (HZ) spectrum.
Remember that in the standard model, where only 
gravity plays a role, the evolution of perturbations may
be studied independently of the exact mechanism responsible
for their generation.   One only needs to postulate an 
initial spectrum, such as a power--law.  In these terms, Inflation 
is just one possible physical model attempting to explain the
origin of the initial power spectrum.  After this
brief discussion, we shall no longer be concerned with
the details of Inflation, and just adopt a power--law
primordial spectrum.

	Once inside the horizon, the perturbations evolve 
differently depending on the cosmic epoch, the various 
physical situations having been discussed in the previous
two sections.  Focus attention
first on the cold dark matter (CDM), which is a cold 
collisionless fluid; hence, no pressure terms.  
The CDM perturbations remain constant after horizon crossing in 
the radiation--dominated epoch, and they only 
begin to grow, on all scales, after matter domination.
This means that, for wavelengths smaller than the equality horizon, 
$k^3P(k)\propto k^{\gamma}$, or $P(k)\propto k^{-3+\gamma}$.  
On the other hand, a perturbation entering the horizon during the 
matter--dominated epoch begins to grow immediately. 
Horizon crossing occurs when $k=2\pi/\tau$, and because
in the matter era $a\propto \tau^2$ and 
$P(k)\propto a^2$ (our perturbation result {\bf IA}),
we deduce that, on scales larger than the equality horizon,
$P(k)\propto k^{1+\gamma}$.
Now we change notation and define the {\em spectral
index} as $n\equiv 1+\gamma$, so that at time $\tau$
\begin{eqnarray*}
P(k) & \propto & k^{n-4} \ \ \ \  \tau_{eq}<\tau<\lambda \\
P(k) & \propto & k^n  \ \ \ \ \ \ \tau_{eq}<\lambda<\tau \\ 
P(k) & \propto & k^{n-4} \ \ \ \  \lambda<\tau_{eq}
\end{eqnarray*}
where $\tau_{eq}$ refers to the comoving horizon at the 
matter--radiation transition (or the equality epoch).   
With $n=1$ the power--law on 
large scales is the more 
familiar form of the HZ spectrum, but its scale invariance is seen 
directly not in the power law, but in the behavior of each 
mode as it crosses the horizon.

	Figure 1 shows several CDM--like power spectra,
where we can see the features just described.  There is
the power--law decrease of $k^3 P(k)$ towards large scales
with the gradual break to a constant towards smaller
scales.  The spectra are plotted as
a function of the mass enclosed in a sphere of 
radius $\lambda$.  This permits us to easily
identify galactic ($M\sim 10^{12} \msun$)
and cluster ($M\sim 10^{15} \msun$) scales.
In these terms, the scale at the break corresponds
to the mass enclosed within the horizon at equality,
namely $M_{eq} = 10^{15} (\Omega h^2)^{-2} \msun$.

	It is of fundamental importance that we have 
discovered a physical scale in the final power 
spectrum.  This scale was not put in by hand; on the
contrary, a featureless power--law was injected
as an initial condition (the simplest general class imaginable).
It is the subsequent evolution that has imposed this scale.
This is extremely important, because we did not 
put in this scale to explain the galaxy distribution
we observe today, but the fact is that this general form
is exactly what is needed to explain both the COBE temperature
perturbations (subject of the next section) and the overall
amplitude of the galaxy fluctuations today.  This is 
a remarkable fact that should not be lost on the reader.
It is often remarked today that the COBE observations have
killed the standard CDM model with $\Omega=1$, $h=0.5$, etc...
The reason for this amounts to a factor of 2: once such 
a standard spectrum has been normalized on the largest 
scales by the COBE observations, the amplitude on galaxy cluster
scales (many orders of magnitude below those of COBE) is 
{\em too high} by a factor of about 2.  This is in fact enough
to put the standard CDM model in jeopardy, but at a first look,
one should be extremely impressed that the general form
of $P(k)$ works so well over such a vast range of scales.
The game today is to slightly change the matter density
to fix this factor of 2.  This works in part because the horizon
scale at equality depends on the non--relativistic matter
content of the Universe.  Lower $\Omega$, with or without a 
cosmological constant, and you delay matter domination,
increasing the scale of the break in $P(k)$.
This is easily seen in Figure 1. 
The implications of a factor
of 2 can be quite large - remarkable overall consistency, but
in detail perhaps requiring a low density parameter.  Not bad!

	Most of the above discussion of the power spectrum
concerns the evolution of the collisionless CDM.  How about
the baryons and photons?  They 
behave as a single fluid dominated by the pressure of the
photons before decoupling.  As each perturbation in this
fluid enters the horizon, it begins to oscillate as
a sound wave, according to our perturbation solutions;
this will become important when discussing the ``Doppler''
peaks of the temperature anisotropy power spectrum 
in the next section.
It is only after decoupling, when the baryons
are no longer hampered by pressure, that 
the perturbations can grow.  Then, with the additional
``pull'' of the already growing CDM perturbations,
the baryons eventually catch--up and develop
a power spectrum similar to that of the dark matter. 
These perturbations then gradually collapse into
dark halos containing baryons and form galaxies
and galaxy clusters.  The photons, 
on the other hand, just simply free--stream through space
after decoupling, for the most part not interacting with 
anything else until the moment some of them are captured 
in one of our CMB experiments.

\section{Temperature Anisotropies}

	The study of the evolution of density perturbations 
in the expanding Universe dates back to well before the prediction or
discovery of the CMB.  Lema\^{\i}tre himself was one 
of the first to propose that structure could be formed by such
perturbations in his primeval atom picture.  He also
eagerly sought the relic radiation of the initial
explosion, thinking that it was probably the cosmic rays.
It was only much later, in 1948, that Gamow and 
his collaborators took things farther in  
suggesting that elemental abundances were established in 
the early Universe during a period of primordial nucleosynthesis
dominated  by {\em thermal} radiation.
They predicted that this radiation should still 
be omnipresent today, with a temperature of about 5K.
Not until quite a bit later was this radiation detected, 
and at first thought to 
be an unexpected noise signal, in Bell Laboratory's
Homdel telescope 
(Penzias \& Wilson 1965; Dicke et al. 1965).  
Peebles (1993) and Partridge (1995) provide 
a history of the discovery of the CMB.  
Soon after this discovery and the associated rise to 
popularity of the Big Bang model, it was realized that the 
background radiation should carry the imprint of 
the density perturbations at early times in the form of 
temperature anisotropies 
(Sachs \& Wolfe 1967; Peebles \& Yu 1970).  
A new and powerful test of the Big Bang scenario
was thus posed, and what was to become a much longer 
than expected search for these CMB anisotropies was begun,
culminating finally in 1992 with their first detection
by the COBE satellite (Smoot et al. 1992).  
Today, there are more than a dozen detections
spanning many angular scales, as shown in Figure 3.
For a good reference to the various experimental 
efforts in progress, see Smoot (1997).

	In this chapter, we investigate the relationship between the 
density perturbations, studied at length in preceding 
section, and CMB anisotropies.  
Our ultimate goal is to explain the general features
of the standard model anisotropies as presented, for example,
by the curves in Figure 3.
This is a subject rich in physical concepts; 
it may be as detailed and complicated
as necessary to make predictions good to arbitrary accuracy, or as 
simple and naive as required to grasp only the essentials.  
In the following, we will search the middle ground, principally by 
developing ideas within the framework of a rather reasonable approximation 
known as the {\em tight--coupling approximation}.   
We shall once again only consider {\bf adiabatic}
perturbations.

  	To understand the nature of this approximation,
recall the state of the Universe at the
decoupling era, the surface we actually ``see'' in a map of the CMB
(exactly what we observe will become more clear as we go along).  
At that time, the Universe consisted of a fluid with several 
components, namely baryons (mainly protons and Helium nuclei), free 
electrons, photons, neutrinos and, perhaps, a form of non--baryonic 
dark matter.  Thomson and Coulomb scattering strongly 
coupled baryons, electrons 
and photons together, ensuring that they behave as a single fluid.  This 
remained the case up til moments just before recombination, 
when the optical depth to scattering begins to become larger and larger.
{\em Recombination} occurs when the majority of the ions
have recombined, but actual decoupling of the photons, when
their mean free path becomes larger than the particle
horizon, occurs slightly afterwards.  For 
standard recombination at redshifts near 1000, the
transition from opacity to transparency is relatively rapid.
The {\em tight--coupling approximation} describes decoupling and
recombination as a single event, an infinitely rapid transition 
before which the baryon--photon fluid existed as a single entity
and after which the baryons and photons are completely decoupled
fluids; in other words, the surface of last scattering is imagined
to be infinitely thin.  In actuality, this is not too crude an 
approximation.  The tight--coupling picture greatly simplifies 
the problem precisely because it permits us to treat
the baryons and photons as a single fluid in the
perturbation equations of the previous section.  
After decoupling, we again
have a simple situation in which the photons simply
free--stream away from the surface of last scattering.

	Sufficiently armed with our tight--coupling approximation, 
we will seek to understand the general form of the
standard model anisotropy power spectrum.  We expect 
that Thompson scattering 
will be irrelevant on large angular scales (small $l$), 
that what we see reflects the initial photon 
perturbations and only purely kinematic gravitational effects, 
because all causal physics is unimportant.   
In fact, on these large scales, the tight--coupling
concept serves no purpose; it is only on sub-horizon
scales at decoupling that this concept has 
a useful meaning.  The anisotropies on 
large scales can all be related to the gravitational field,
and we call the relevant physics the Sachs--Wolfe (SW) effect.  
This is one of the two primary physical effects we need
to understand in order to construct the power spectrum,
and it will be treated in section 5.3.
Although one often hears that
the Sachs--Wolfe effect is only gravitational, the actual 
contribution to the observed anisotropies from the
initial photon distribution and from gravitational 
redshift is {\em gauge dependent}, i.e., the individual 
contributions depend on the particular gauge chosen for the 
calculation.  The final result is all the same gauge 
invariant, because it is what we actually observe. 
On angular scales below the decoupling horizon (large $l$), casual
physics, i.e., pressure effects,  will govern the evolution 
of the perturbations,
and hence the production of anisotropies.  
This is the other key physical effect we need to construct
the power spectrum, and it is treated first, in section 5.2.
In the tight--coupling limit, the photons and baryons 
act as a fluid with large pressure, and  
we have seen that sub--horizon perturbations in 
such a fluid become sound
waves.  We expect to see their signature in the
CMB anisotropies on smaller angular scales; 
they produce the famous ``Doppler peaks'' predicted
around degree angular scales, demonstrated by the 
curves in Figure 3.

	We saw before how the horizon size at the 
matter--radiation transition became the only important scale 
characterizing the final density perturbation power spectrum. 
From this discussion, we should expect to find that the angular
size of the horizon at decoupling ($\sim 0^o.5$
$\Omega^{1/2}$) will impose itself as a founding scale
in the CMB anisotropy power spectrum.  This is an important
point; it will, among other things, offer us a way to
constrain the global geometry of the Universe.	

\subsection{Describing the CMB sky}

	Before we can dive into calculations of 
CMB anisotropies, we must decide {\em what} to calculate.
This depends on the quantities used to describe the
anisotropies.  Just as for the density field, we
describe the CMB sky as a {\em random field}, in 
two dimensions: $\Delta(\n)\equiv \delta T(\n)/T$, 
where $\n$ is a unit vector on the sphere.  
This means that the value of $\Delta$ at each
position on the sphere (sky) is a random variable.
By definition $<\Delta(\n)>=0$, and the
two--point (auto--) correlation function gives 
the covariance of these random variables as
\begin{displaymath}
C(\theta) = <\Delta(\n_1)\Delta(\n_2)>
\end{displaymath}
with $\cos\theta=\n_1\cdot\n_2$.  
Notice that the correlation function only
depends on the separation of the two 
sky directions, $\theta$.  This is once
again the work of the cosmological principle:
the nature of the anisotropies must be 
independent of position or orientation 
on the sky.
This is a statement that the temperature
fluctuations were generated by a 
{\em statistically} isotropic and homogeneous process.
Because the mean of $\Delta$ is by 
construction zero, Gaussian anisotropies, as 
in the standard model,
are completely specified by the 
two--point correlation function.

	It once again proves convenient to 
work with the harmonic modes of our random field.
On the sphere we employ the spherical harmonic transform:
\begin{displaymath}
\Delta(\n) = \sum_{lm} a_{lm} Y_{lm}(\n)
\end{displaymath}
The coefficients $a_{lm}$ are 
random variables with $<a_{lm}>=0$.  
We should expect, as before for the density field, 
that their covariance is related to the two--point
correlation function.  To find this
covariance, first note that since
the two--point correlation
function only depends on separation,
we may expand it in a Legendre series
\begin{eqnarray}
C(\theta) & = & \frac{1}{4\pi} \sum_l (2l+1) C_l P_l(\n_1\cdot\n_2)\\
	  & = & \sum_{lm} C_l Y_{lm}(\n_1) Y^*_{lm}(\n_2)
\end{eqnarray}
where we have used the addition theorem of spherical harmonics
to get to the second line.  This permits
us to easily calculate the quantity
\begin{eqnarray*}
<a_{lm}a^*_{l'm'}> & = & \int\int d\Omega_1 d\Omega_2 \;
	<\Delta(\n_1)\Delta(\n_2)> Y^*_{lm}(\n_1) Y_{l'm'}(\n_2)\\
	& = & \int\int d\Omega_1 d\Omega_2 \;	
	C(\theta) Y^*_{lm}(\n_1) Y_{l'm'}(\n_2) 
\end{eqnarray*}
with the important result
\begin{equation} 
<a_{lm}a^*_{l'm'}> =  C_l\delta_{ll'}\delta_{mm'}
\end{equation}
The presence of the $\delta$ functions 
tells us that the $a_{lm}$ are independent random
variables.  This is a direct consequence of the
fact that the two--point correlation function only
depends on separation, i.e., the cosmological
principle.  The set of $C_l$ is the (angular) {\em power
spectrum} of the temperature anisotropies, and 
for the Gaussian fluctuations encountered in the 
standard model, it is a complete description of 
the CMB sky, just as the density power spectrum
is a complete description of the density field 
in Gaussian models.  {\em The angular power spectrum is thus 
the fundamental quantity} in the theory of
CMB anisotropies of the standard model.  Non--Gaussian
theories would require consideration of higher 
order moments of the field $\Delta$.  

	Our goal is now clearly before us: to find
the $C_l$ and their relation to $P(k)$ of the
density field.  Specifically, we wish  
to construct and understand the overall form 
of the angular power spectrum of the standard model
anisotropies, as for example presented in Figure 3.
This work is the analog of our efforts to understand
the general form of the density field as described by
$P(k)$. 

\subsection{Start with the simple}

	So, just how do the density perturbations 
leave their imprint on the photons?  We
proceed by first ignoring various effects, to be
re--added later.  Suppose, then, that
in addition to our basic assumption that the surface
of last scattering is infinitely thin, 
space is flat and static (no curvature, no expansion)
and completely uniform and homogeneous after decoupling.
This means in particular that, although there are 
perturbations on the last scattering surface, they
do not exist between us and this surface, and that
there are no important photon interactions off the
surface of last scattering;
this is all, of course, highly unrealistic,
but we will remedy that as we go along.  We solve
this problem, a standard problem of classical physics,
in this subsection.  This will give us some
understanding of how we directly ``see'' the photon 
perturbations on the last scattering surface.
The intuition we gain in this manner will in practice
only apply to small angular scales.  To 
go beyond this, we will need to 
introduce some important concepts of {\em photon
transport} in GR, presented in the following subsection.
These will permit us to add back into the problem the 
expansion and curvature of space, as well as perturbations 
after decoupling.  It is here that we encounter the famous 
Sachs--Wolfe (SW) effect, which dominates the CMB
anisotropies on large angular scales.

\begin{figure}\label{fig:diag1}
\begin{center}
\includegraphics*[height=16cm,width=13cm,angle=-90]{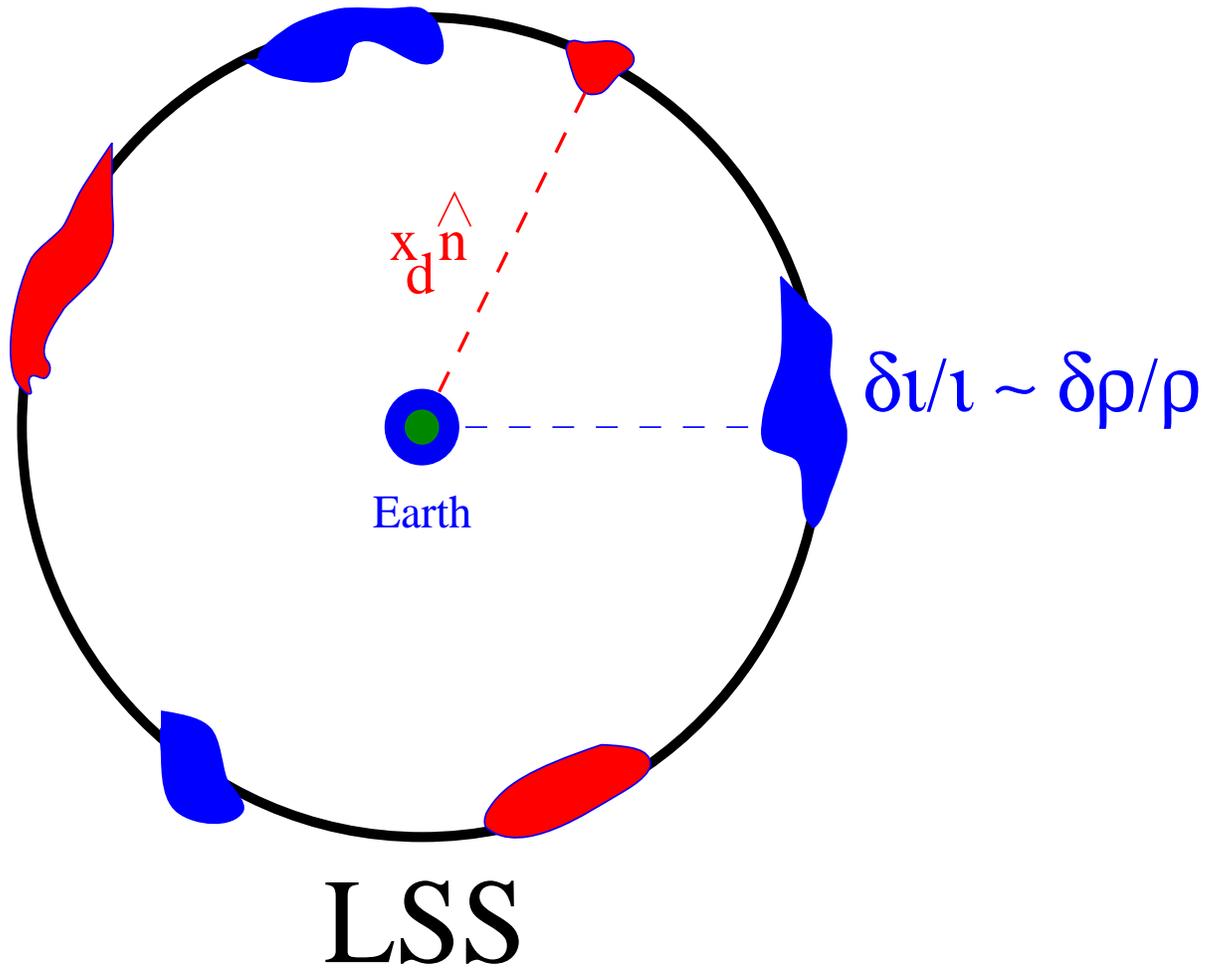}
\end{center}
\caption{Schematic diagram of the last scattering surface.
The observer is at the center and sees the fluctuations
on the last scattering surface at a distance of 
$x_d\n$.  In the tight--coupling approximation,
the photon intensity perturbations on the surface, 
$\delta\ii/\ii$, are equal to the photon density 
perturbations, $\delta \rho_\gamma/\rho_\gamma$.
On sub--horizon scales, the baryon--photon 
fluid oscillates with sound waves, which are then
seen by the observer as peaks and valleys in the
$C_l$ verses $l$ plane (Eq. \ref{eq:Tanis:Clint}).}
\end{figure}

	We have before us a straightforward
problem of classical physics: how to describe 
the intensity pattern on a distant screen (see Figure 2).  More
precisely, we place ourselves at the origin of a 
sphere and the screen corresponds to the inside 
surface of this sphere at a distance of $x_d$
(the distance to the last scattering surface).
Suppose that the intensity pattern on the screen is given by
$\ii(\hat{n},t_d)$, where $t_d$ is the time at
decoupling, when the light is emitted from the
screen.  The intensity that we measure today, at time $t_o$ sitting 
at the origin of our coordinate system ($\vec{x}=0$), and coming from
direction $\hat{n}$, is $\ii(-\n,t_o,\vec{0}) = 
\ii(-\n,t_d,x_d\n)$,
because surface brightness is conserved
(remember, we are ignoring expansion for the time being).
In the tight--coupling limit, the radiation intensity at
decoupling may be 
written in terms of the photon energy density, $\rho_\gamma$, as
$\ii = (c/4\pi) \rho_\gamma$.  Notice that this means that
the photon intensity at an arbitrary point in the baryon--photon
fluid is independent of propagation direction; this is a 
direct consequence of the tight--coupling assumption, and
the breakdown of this exact condition is one of the 
complications to be treated in a more complete calculation.
Thus, we have arrived at the statement 
\begin{eqnarray}
\Delta  & \equiv & \frac{\delta \ii}{\ii}(-\n,to,\vec{0}) = 
	\delta_\gamma(x_d\n,t_d) \\
\label{eq:Tanis:Delta}
& = & \frac{1}{(2\pi)^3} \int d^3k\; e^{-ix_d\n\cdot\vec{k}}
	\delta_\gamma(\vec{k},t_d)
\end{eqnarray}
In the last line we have expressed the photon
perturbation in terms of its Fourier transform, the 
kind of quantity we used when discussing density
perturbations previously.  Using an identity which 
translates basis functions from Cartesian coordinates to spherical
coordinates,
\begin{equation}
e^{-i\vec{k}\cdot\vec{x}} = 4\pi \sum_{lm}\; (-i)^l j_l(kx)
	Y^*_{lm}(\hat{k}) Y_{lm}(\hat{x})
\end{equation}
we can write the observed intensity perturbation today
as 
\begin{equation}
\Delta(\hat{n}) = \sum_{lm}\; \left[ 4\pi (-i)^l \frac{1}{(2\pi)^3}
	\int d^3k\; j_l(kx_d) Y^*_{lm}(\hat{k}) 
	\delta_\gamma(\vec{k},t_d) \right] Y_{lm}(\hat{n})
\end{equation}
from which we learn that the $a_{lm}$ are given by the
expression in square brackets; this is the sought
relation between the perturbations and the $a_{lm}$
describing the observed anisotropies of the CMB.
Not so difficult in the end!

	It is clear from the linear nature of this 
relationship that the $a_{lm}$ retain all the 
statistical properties of the perturbation variables;
if these latter are Gaussian random variables, as in
the case of inflationary models, then so are the
spherical harmonic coefficients.  We can be more
quantitative.  It is easy to see that $<a_{lm}> = 0$,
which is no surprise.  As for the {\em covariance} of
the harmonic coefficients, this is calculated as 
follows:
\begin{eqnarray}
<a_{lm} a^*_{l'm'}> & = & \frac{1}{4\pi^4} (-i)^l (+i)^{l'}
	\int d^3k\; d^3k' j_l(kx_d) j_l(k'x_d) \\
& & 	Y^*_{lm}(\hat{k})Y_{l'm'}(\hat{k}')<\delta_\gamma(\vec{k},t_d)
	\delta_\gamma(\vec{k}',t_d)>
\end{eqnarray}
By using Eq. (\ref{eq:lls:deltacov}) and recalling
that $<a_{lm} a^*_{l'm'}> = C_l \delta_{ll'} \delta_{mm'}$, we find
an expression for the $C_l$:
\begin{equation}\label{eq:Tanis:Clint}
C_l = \frac{2}{\pi} \int dk\; k^2 P_\gamma(k) j^2_l(kx_d)
\end{equation}
Within the limits of our simplifying assumptions,
which we will now gradually relax, this is a
central relation between CMB anisotropies
and density perturbations at the epoch of last
scattering.  If the latter are calculated
using perturbation theory, and note that here
we specifically need the {\em photon} perturbation,
$\delta_\gamma$, then we can find the $C_l$ that
an observer would deduce.  In the tight--coupling
approximation, $\delta_\gamma$ is easily calculated from the
evolution of the baryon--photon fluid.  Corrections to the this 
approximation render things more complicated 
precisely because we are saying that the two 
components do not behave in exactly the same
manner.  It is here that one must start to 
employ the Boltzmann equation to perform an
accurate calculation.  

	We have achieved one of our goals, that of
describing the the relationship between the
density perturbations and the anisotropies on 
small scales.  In the tight--coupling limit,
the baryon--photon fluid is oscillating on
the last scattering surface, creating
the Doppler peaks seen in the $C_l$.
We have indeed made
many assumptions to arrive at this result, but
the important point is that the essence of the
physics of the problem is revealed without
excessive detail (to be added at will!); and
the line of reasoning leading to these results
will be reused in the following.

\subsection{A traveling photon}

	We wish to relax the simplifying restrictions
employed in the last subsection.  Among other things,
we are now forced to consider more carefully just 
exactly how a photon travels between us and the
last scattering surface.  This is a {\bf must}
if we wish to understand the influence of 
perturbations after decoupling.  Consider a
packet of photons, or light ray, following 
a trajectory $x^\alpha(\lambda)$ in spacetime and
described by four momentum $p^\alpha(\lambda)$.  
The {\em affine parameter} $\lambda$ labels
position along the photon path.  Photons
are zero mass particles, so they travel
along {\em null geodesics}:
\begin{eqnarray}\label{eq:PT:ng}
\frac{dp^\alpha}{d\lambda} &  + & \Gamma^\alpha_{\beta \gamma} 
	p^\beta p^\gamma = 0\\
\nonumber
 & & p^\alpha \equiv \frac{dx^\alpha}{d\lambda} \\
\nonumber 
& & p^2 = 0
\end{eqnarray}
The affine parameter is {\em defined} by the second
equation; the third specifies that the path is a null
geodesic.  We already see from this basic
equation how gravity influences the CMB --
via the Christoffel symbols characterizing
the spacetime geodesics.
This is in fact the equation we will employ to 
find $\delta T/T$, thanks to the fact that 
Gravity is color blind; however, it is useful to at
least see a little of the Boltzmann equation,
which is used in more complete calculations
of temperature anisotropies.  

	Evolution of the photon ``gas'' can be
modeled by its phase space distribution
function, $f(x^\alpha,p^i)$, and we assume that
this gives a complete description of the
CMB (no higher order correlations or
collective effects in the gas).  It is important
to notice that $f$ here is taken to be a function
of the contravariant four--position and of the 
{\em contravariant three--momentum}.  This is not
obvious, because it is in fact the covariant
three--momentum which is the dynamically conjugate
variable to position, but the isomorphism between
covariant and contravariant vectors 
($p^\alpha = g^{\alpha\beta} p_\beta$) permits us to
do this.  Since we will be primarily interested in the
free streaming of CMB photons after scatterings have 
ceased to be important (tight coupling
means that we can forget about photon transport
before decoupling, treating everything as a simple
fluid), we employ the Liouville equation (Boltzmann
equation with no sources)
\begin{displaymath}
\frac{Df}{D\lambda} = p^0\frac{\partial f}{\partial x^0} + 
	p^i\frac{\partial f}{\partial x^i}
	+ \frac{d p^i}{d\lambda}\frac{\partial f}{\partial p^i} = 0
\end{displaymath}
Gravity enters here through the geodesic equation 
describing the photon path, Eq. (\ref{eq:PT:ng}).

	How is this equation actually used?  The first
thing one must do in a general situation is to find
the form of the null geodesics, $x^\alpha(\lambda)$ and 
$p^\alpha(\lambda)$, given a coordinate system [Christoffel 
symbols $\Gamma^\alpha_{\beta \gamma}(\lambda)$].  
These quantities are then plugged into the 
Liouville equation, which one can now integrate
to find the solution $f[x^\alpha(\lambda),p^i(\lambda)]$.
We do this, as an illustration, for the perfectly
homogeneous and flat FRW background.  This is only
for illustration; as mentioned, in our discussion
on perturbations we shall in fact get by with only
the geodesic equation.  The Christoffel symbols 
in conformal coordinates ($\tau, \vec{x}$) are given by Eqs. 
(\ref{eq:rcosmo:gamb})--(\ref{eq:rcosmo:game}), 
with all perturbation variables set to zero. 
With flat space--like slices,
we know that all geodesics are ``normal'' straight lines,
i.e., we can describe them by $x^1(\lambda)=x^0(\lambda)=\tau(\lambda)$,
$x^2=x^3=0$ and $p^1=p^0\equiv p(\lambda)$, $p^2=p^3=0$ (notice that 
$p^2=0$ as required).  The zero component of the
geodesic equation then yields $p\propto 1/a^2$, which you
find very upsetting, because everybody knows that
the redshift is $\propto 1/a$.  So what happened?!
Everything is in fact OK, because this momentum is the
conjugate variable to $\tau$, i.e., $p=d\tau/d\lambda$;
the conjugate variable to cosmic (local observer) time $t$,
let's call it $q$, is related via $q=dt/d\lambda
=(d\tau/d\lambda)(dt/d\tau) = a p \propto 1/a$, and
we are reassured.  Moral: let's not forget that we
are manipulating conformal time in the following.
The space components of the geodesic equation simply
confirm the fact that $p^1=p^0=p$ and $p^2=p^3=0$ all
along the trajectory, that our choice is consistent.

	Now put these results into the Liouville equation
to find
\begin{displaymath}
p\frac{\partial f}{\partial\tau} + p\frac{\partial f}{\partial x^1}
	+ \frac{dp}{d\lambda}\frac{\partial f}{\partial p} = 0
\end{displaymath}
or
\begin{displaymath}
 \frac{\partial f}{\partial\tau} + \frac{\partial f}{\partial x^1}
	- 2\frac{\dot{a}}{a} p \frac{\partial f}{\partial p} = 0
\end{displaymath}
In our perfectly uniform background, we also know that
$f$ does not depend {\em explicitly} on position, $x^i$,
so we are left with a very simple expression:
\begin{displaymath}
\frac{\partial f}{\partial\tau} - 2\frac{\dot{a}}{a} p 
	\frac{\partial f}{\partial p} = 0
\end{displaymath}
whose solution is $f(\tau,p) = f(\tau_i,pa^2/a_i^2)$,
where $\tau_i$ is some initial time at which the
phase function is given (this is
enough, because the Liouville equation is first order).

	Suppose that at some initial time (in a space--like
slice at constant $t$ or $\tau$) all comoving
observers see the Universe filled with blackbody radiation.   
Everyone thus remarks that $f = (e^{\omega/T}-1)^{-1}$, 
where $\omega$ is the observed frequency, or energy, as measured 
by a comoving observer.  We must be very careful 
to correctly interpret our momentum variable in 
terms of this observed energy, the main 
difficulty being that we are working in a 
gauge where $g_{00}\ne 1$.  In general, the
energy measured by an observer with 4--velocity
$u^\alpha$ is given by the covariant expression
$\omega = p^\alpha u_\alpha$.  Using the condition
$u^\alpha u_\alpha = g_{\alpha \beta}u^\alpha u^\beta = 1$
and noting that comoving observers have $u^i=0$,
we see that $u^0 = g_{00}^{-1/2}$; hence, $\omega = g_{00}p^0u^0
= g_{00}^{1/2}p^0 = ap$, and {\bf not} simply $p^0$. 
Thus, $f(\tau_i,p) = (e^{a_ip/T}-1)^{-1} \equiv B(a_ip/T)$, and
this is the same for everybody.  
Our result is that at later times
$f(\tau,p) = B[a^2p/(a_iT)] = B[\omega/(a_iT/a)]$, from which one 
concludes that the spectrum is still thermal, but now
with a temperature $Ta_i/a$.  This is the familiar redshift 
law for the CMB temperature.  The result is immediately
obvious from the geodesic equation, which tells
us that all photons experience the same redshift, but
this derivation helps to understand the workings of 
the Liouville equation.

	Let's now turn our attention to the effects
of metric perturbations on the CMB.  Fortunately,
gravity is color blind, meaning that it does the
same thing to photons of any frequency (this is 
a consequence of the equivalence principle).  Thus,
metric perturbations will not alter the thermal nature
of the CMB frequency spectrum.  For this reason, we can
get quite far in our analysis by only using the
geodesic equation for the photons; whatever happens
to one set of photons happens to all of them following
a particular path, and so this evolution must therefore 
also be the same as the temperature of the thermal 
spectrum.  We choose to work in {\bf Poisson gauge} and, to keep 
things as simple as possible, with a flat space background.  
Only scalar perturbations in an ideal fluid
will be considered, so 
the only relevant metric perturbation variable is the
``Newtonian'' potential, $\psi$.  The technique
is to expand the geodesic equation (\ref{eq:PT:ng}) out to first order
in the perturbation variables.  
Our primary concern is with the time
component
\begin{equation}\label{eq:PT:pertgeo}
\frac{dp^0}{d\tau} =  -\frac{\dot{a}}{a}\left( p^0 +
	\gamma_{ij}p^i\frac{p^j}{p^0}\right) 
	+ \dot{\psi}\left(\gamma_{ij}p^i\frac{p^j}{p^0} -p^0\right)
	+ 4\frac{\dot{a}}{a}\gamma_{ij}\frac{p^ip^j}{p^0}\psi 
	- 2p^i\psi_{,i} 
\end{equation}
Notice that the original derivative with respect to $\lambda$ 
has been converted to a derivative with respect to $\tau$ (by
dividing through by $p^0$).  Imposed on this 
equation is the condition $p^2=0$:
\begin{equation}\label{eq:PT:zeromass}
(1+2\psi)(p^0)^2 - (1-2\psi)\gamma_{ij}p^ip^j = 0
\end{equation}
from which we immediately learn that the expression
in parenthesis in the second term of Eq. (\ref{eq:PT:pertgeo})
is first order, and hence the whole second term may be
dropped.  By defining $p\equiv \sqrt{\gamma_{ij}p^ip^j}$,
we put Eq. (\ref{eq:PT:zeromass}) into the compact form
\begin{displaymath}
p^0 = (1-2\psi)p
\end{displaymath}
We may then also write $p^i = p n^i$, where the $n^i$ are
the direction cosines along the photon path such that
$\gamma_{ij} n^i n^j =1$.  To first order, this reduces the geodesic
equation to
\begin{displaymath}
\frac{1}{a^2p^0}\frac{d}{d\tau}(a^2p^0) = -2 n^i\psi_{,i}
\end{displaymath} 
where we have used the fact that $dx^i/d\tau = p^i/p^0$
is equal to $n^i$ {\em to zero order}, as appropriate 
to keep the final expression to first order. 

	We now use some of our previous results concerning the
evolution of the gravitational potential, $\psi$.  In the
matter--dominated era, when $x\sim c_s^2 \sim 0$, we have
seen that $\dot{\psi}=0$ for modes outside the horizon,
$\lambda>\tau$.  The same holds for sub--horizon perturbations
when the curvature/cosmological constant terms are unimportant,
i.e., at all times in a critical model or at sufficiently early 
times in non--critical models.  This is most
easily seen from our Newtonian result that $\delta_g\propto a$;
hence, the Poisson equation tells us that $\phi_k \propto \rho_b a^2 
\delta_g = const$ (matter--dominated epoch).  For the following
calculation, we shall in practice adopt the critical model
and ignore any possible residual effects of the
radiation--matter transition at the moment of last 
scattering; we will make some
brief comments below concerning modifications in non--critical
models (in passing, we note that even in flat models ($\Lambda\ne 0$)
the radiation--matter transition approaches the last scattering
surface as $\Omega$ is lowered).  Then, we may write
\begin{displaymath}
\frac{d\psi}{d\tau} = \dot{\psi} + n^i\psi_{,i} = n^i\psi_{,i}
\end{displaymath}
leading to our final result
\begin{equation}\label{eq:rcosmo:pgeo}
\frac{1}{a^2p^0}\frac{d}{d\tau}(a^2p^0) = -2 \frac{d\psi}{d\tau}
\end{equation}
	 
	Apart from factors of $a$, this is the traditional result
for the gravitational redshift due to a potential $\psi$; 
and no, the factor of 2 is not a typo!  The rational for 
this factor of 2 provides a good exercise in understanding measurements
in GR.  Key question in all situations: what are we trying to 
calculate, or what is the measurement that interests us?
We really want to find the {\em frequency} (energy) shift 
as measured by two observers, one at each end of the photon 
path and each at rest relative to the coordinate system.
As already mentioned, the fact that we are working in a gauge 
(coordinate system) where $g_{00}\neq 1$ means that
we must use extreme care in 
interpreting $p^0$.  We have two observers, one at the 
emission point, $x_e$, and the other at the reception 
point, $x_r$.  Since each is at rest relative to the
coordinate frame in use, each one has a four--velocity with
zero spatial components.  From the requirement that 
$u^\alpha u_\alpha=1$, we find, as before, that $u^0=g_{00}^{-1/2}$.
This time, working to first order, we thus obtain 
the measured frequency at each point 
$\omega=g_{00}p^0u^0=a(1+\psi)p^0$.
Consider first the classic case encountered in most textbooks,
i.e., $a=1$, a constant which may be dropped from our 
expressions.  To first order, the geodesic equation
then yields a frequency shift of
\begin{displaymath}
\frac{\Delta\omega}{\omega} = -2\Delta\psi + \Delta\psi = -\Delta\psi
\end{displaymath}
a reassuring and familiar expression.  

	How about when $a$ is not constant, as in our 
cosmological setting?  Let us again be very careful
and explicit about what it is we should calculate in
our perturbation theory.  We want the temperature
perturbation induced by $\psi$, {\em relative to the
unperturbed background model}.  In the background model
(pure FRW), we know that $a\omega=const$, i.e., $\Delta(a\omega)=0$.
Thus we should find $\Delta(a\omega)/(a\omega)$ -- this is the
$\delta T/T$ that we seek. Because the numerator of this
ratio is of first order, the denominator is left to
order zero -- a constant -- and so may be specified at 
either end of the photon path. In the following
expression, $r$ and $e$ refer to the points of reception
and emission, respectively:
\begin{eqnarray*}
\frac{\delta T}{T} & = & \frac{\Delta(a\omega)}{a\omega} \\
	& = & \frac{a_r^2[1+\psi(r)]p^0(r)-a_e^2[1+\psi(e)]p^0(e)}
	{a^2p^0} \\
	& = & \frac{\Delta(a^2p^0) + a^2p^0\Delta\psi}{a^2p^0}\\
	& = & -\Delta \psi
\end{eqnarray*}  
where we have used Eq. (\ref{eq:rcosmo:pgeo}) to get to the
last line.  From the second to the third, we
used the fact that $a^2p^0=const$ to zero order.
The beauty of the result is that it is exactly what
we would have guessed, given the concept that $\psi$
is a kind of Newtonian potential.

\begin{figure}\label{Cl}
\begin{center}
\includegraphics*[height=14cm,width=14cm,angle=-90]{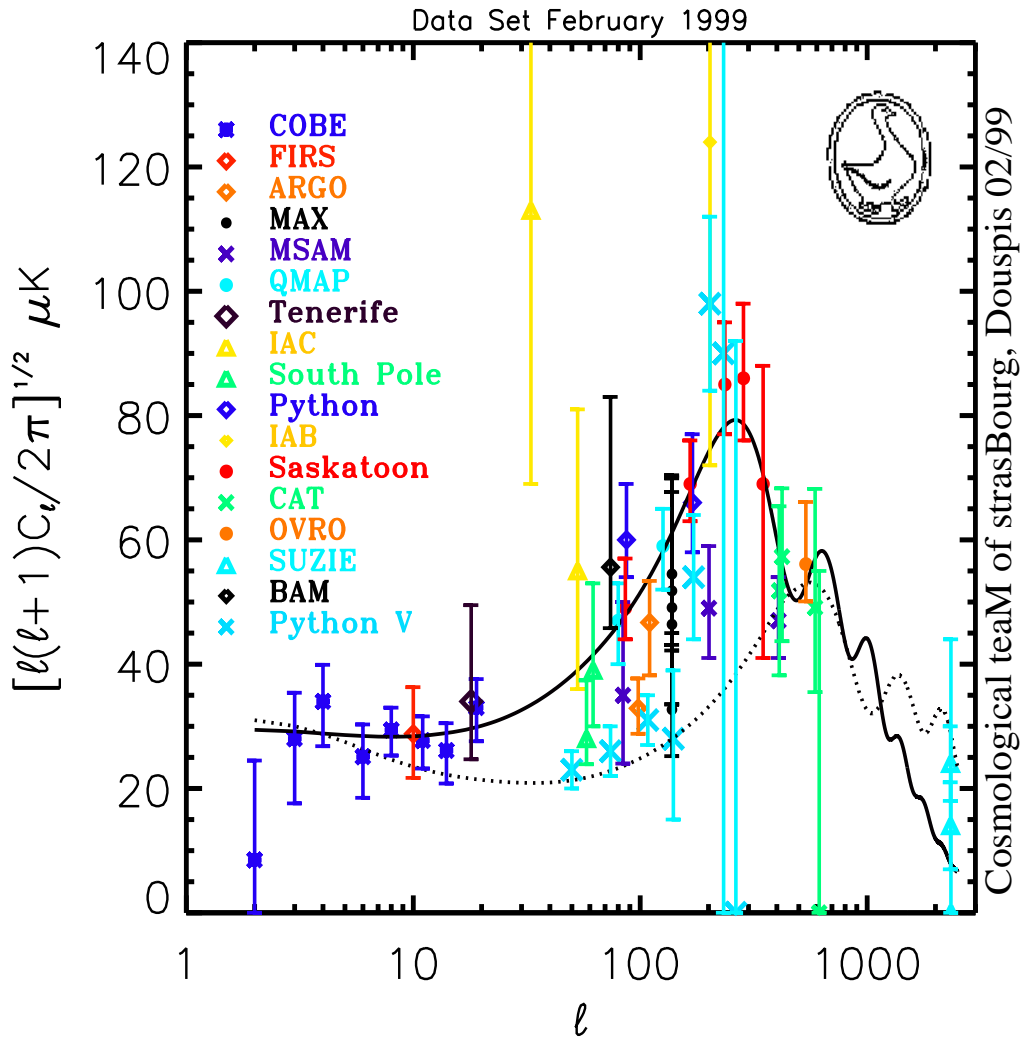}
\end{center}
\caption{Angular power spectrum of the CMB anisotropies.
The data come from the various experiments as indicated by
the color code on the left.  The solid curve is the prediction
for a flat CDM model with $\Omega=0.4$, $\Lambda=0.6$, $H_o=40$ km/s/Mpc, 
$Q=19\;\mu$K, $\Omega_b h^2 = 0.006$ and $n=0.94$, while
the dotted line represents an open model with $\Omega=0.2$,
$\Lambda=0$, $H_o=60$ km/s/Mpc, $Q=20\; \mu$K, $\Omega_bh^2=0.015$
and $n=1$ (no gravitational waves, no reionization).  Late--time
evolution of the gravitational potential in the flat model
is not important enough to alter the basic result expressed 
in Eq. (\ref{eq:Tanis:ClSW}) for $n=1$ -- we see a plateau
in $l(l+1)C_l$.  In the open model, on the other hand,
late--time evolution of $\psi$ is sufficiently important 
to substantially modify the plateau.  
The Doppler peaks at higher $l$ are the projected acoustic oscillations
of the baryon--photon fluid on the last scattering surface, Eq. 
(\ref{eq:Tanis:Clint}).  Curvature in the open model
has shifted the angular position of the first peak to higher
$l$ compared to the flat scenario.  Figure prepared by M. Douspis; see
{\tt http://astro.u-strasbg.fr/Obs/COSMO/CMB/}}
\end{figure}

	To get the classical SW expression, we need now to consider
the initial conditions corresponding to {\bf adiabatic}
perturbations.  Since the relative particle densities
remain constant in this kind of perturbation, we 
may relate the dominate (non--relativistic) dark 
matter perturbation to the CMB temperature perturbation by 
$\delta \propto 3 \delta T/T$.  For super-horizon
perturbations ($\dot{\psi}=0$) on the last scattering surface 
(i.e., on angular scales larger than $\sim 0^o.5$), our perturbation
Eq. (\ref{eq:rcosmo:lspert1}) tells us that, in the initial hyper-surface,
$\delta T/T(e) = (1/3)\delta(e) = -(2/3)\psi(e)$.  We 
remark in passing that in our perturbation theory 
the initial hyper-surface is {\em defined by $\tau=const$};
and this implies that, to first order, $\psi(e)$ is to 
be taken in the surface defined to zero order, which
is the {\em same surface as in the background}.
This leads to
\begin{displaymath}
\frac{\delta T}{T} = -\Delta \psi + \frac{\delta T(e)}{T}
	= -\psi(r) + \frac{1}{3}\psi(e)
\end{displaymath}
Since $\psi(r)$ is a constant independent of direction on
the sky, and what we measure is a difference between different
directions, our answer is the well known result
\begin{equation}\label{eq:PT:SW}
\frac{\delta T}{T} = \frac{1}{3}\psi(e)
\end{equation}
which was the remaining goal we had set before us at 
the beginning of this chapter.  

	To finally get to the $C_l$ plot, we need to translate
the SW relation (\ref{eq:PT:SW}) into $C_l$.  
Remember, the SW effect dominates the temperature
fluctuations on large angular scales, which means
at small $l$.  Using the 
same reasoning as before, in going from Eq. (\ref{eq:Tanis:Delta})
to Eq. (\ref{eq:Tanis:Clint}), we obtain
\begin{equation}
C_l = \frac{2}{3\pi}\int dk\; k^2 P_\psi(k) j_l^2(kx_d)
\end{equation}
where $<\psi_{\vec{k}}\psi^*_{\vec{k'}}> = (2\pi)^3 P_\psi(k)
\delta(\vec{k}-\vec{k'})$ defines the gravitational potential
power spectrum.  Recalling that on super-horizon scales
$P_\psi(k) \propto k^{n-4}$, we see that
\begin{eqnarray}\label{eq:Tanis:ClSW}
\nonumber
C_l^{SW} & \propto & \int dk\;  k^{n-2}
	j^2_l(kx_d)\\
\nonumber
& \propto & \frac{\Gamma(3-n) \Gamma\left(\frac{n+2l-1}{2}\right)}
	{\Gamma^2\left(\frac{4-n}{2}\right)
	\Gamma\left(\frac{5+2l-n}{2} \right)}\\
& \propto & \frac{1}{l(l+1)} \ \ \ \ \ \ \ \ \ \ \ \ \ (n=1)
\end{eqnarray}
Note that the last expression is only valid for the HZ spectrum, $n=1$.
Two examples of CMB angular power spectra are shown in Figure
3 superimposed on the current data set.  The flat model
(solid curve) displays characteristics very similar
to those of a purely critical model; in particular,
as $n\sim 1$ in this model, we see the flat
{\em SW plateau} expected according to Eq. (\ref{eq:Tanis:ClSW}).

	The critical element in this derivation was the
constancy of the gravitational potential $\psi$ along
the photon path, something which breaks down when
the Universe is not completely matter dominated.
This occurs, for example, at late times in non--critical
models and on the last scattering surface if
$\Omega$ is low, because the radiation--matter
transition is moved closer to this surface.  An evolving $\psi$
creates an additional effect often refered to as
the {\em integrated Sachs--Wolfe} (ISW) effect.
This occurs because a photon does not climb out of
the same potential well as it fell into, due to
the evolution of $\psi$.  This adds to the anisotropy 
calculated above and can significantly change
the shape of the spectrum, most markedly at
low $l$.  We see this for the open model shown
in Figure 3: even though the primordial spectrum
has $n=1$, a flat SW plateau no longer exists.
In the flat model shown, the ISW effect is
small enough that we still clearly see a flat
SW plateau similar to that expected from a
purely critical model with $n=1$.  We shall not
go into further detail on this subject. 

\begin{figure}\label{fig:diag2}
\begin{center}
\includegraphics*[height=14cm,width=13cm,angle=-90]{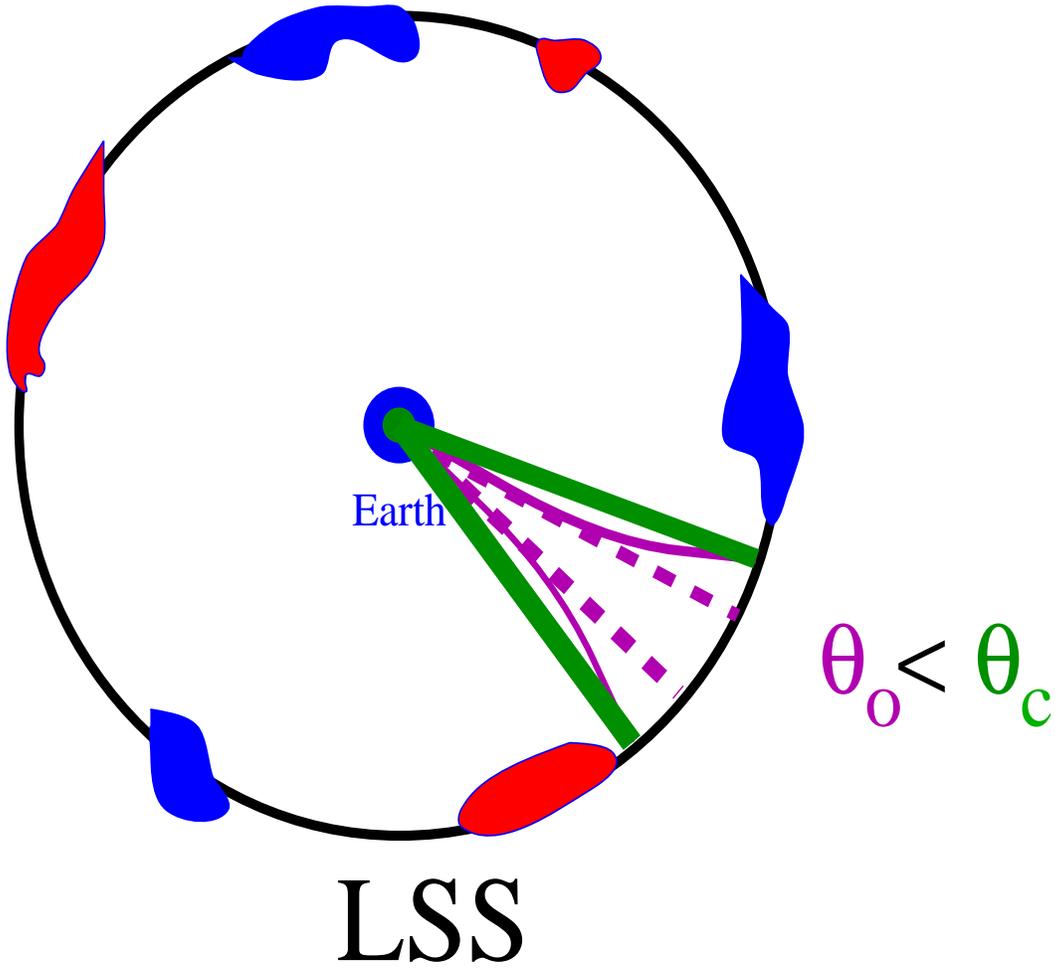}
\end{center}
\caption{Same as Figure 2, but showing null geodesics
in a flat (straight, green lines) and an open (negative 
curvature) Universe (curved, red lines).  The curving
geodesics in the open model project back onto the
surface of last scattering (dashed, red lines) 
with a smaller angle than for the straight geodesics.
The focusing of light rays in a curved Universe 
thus reduces the angular size subtended by the horizon
at last scattering in the open model, $\theta_o$, compared
to the critical model, $\theta_c$.}
\end{figure}

\subsection{Constructing the anisotropy power spectrum}

	We have now in hand the two main results needed 
to understand the overall form of the anisotropy 
power spectrum: on large scales, those larger than
the horizon at decoupling, it is the SW effect which
is responsible for the temperature fluctuations, Eq. 
(\ref{eq:Tanis:ClSW}), while on scales below this
horizon, we observe the acoustic oscillations via
the relation (\ref{eq:Tanis:Clint}); additional anisotropy sources exist
in non--critical models, such as the
ISW effect.  Spectrum plots, such as the one
shown in Figure 3, are made with $l(l+1)C_l$, so that a scale
invariant HZ spectrum (in a critical Universe) 
appears as a horizontal line at low $l$.  Once again,
the horizon is the scale imposed on the problem.
For standard recombination at $z\sim 1000$, the projected 
angular size of the horizon at last scattering is 
$\sim 0^o.5 \Omega^{+1/2}$ (assuming a zero cosmological 
constant).  This angular scale represents a transition from 
the large--scale behavior, e.g., the flat SW spectrum 
of Eq. (\ref{eq:Tanis:ClSW}) in critical models,
to the oscillations of Eq. (\ref{eq:Tanis:Clint}), 
which are seen as the famous ``Doppler'' peaks in the 
anisotropy power spectrum.  We have thus accomplished
our task of explaining the overall form of plots such as 
Figure 3.

	We have not discussed the ISW effect and its
consequences in any detail, nor shall we.  Another
consideration which we only touch on now
in passing concerns the effect
and spatial curvature.  This is extremely important,
because it determines the angular size subtended by
the decoupling horizon, as diagramed in Figure 4. 
The focusing of light rays in a negatively
curved Universe means that the projected angular
size of the horizon at decoupling, as seen by us
today, is smaller than for a critical Universe.  
The position of the first Doppler peak thereby
offers a measure of the curvature of the Universe:
in a model with large curvature, we expect the 
first peak to appear at higher $l$ than in a critical
Universe, as shown by the dotted curve in Figure 3.
These remarks in fact only apply to {\em adiabatic} 
perturbations; the location of 
the first peak is slightly different for isocurvature 
perturbations.  As our goal is not to expose all possible
effects and their dependence on the various cosmological
parameters, we shall end our discussion here, content
with our rapidly acquired understanding of the general form of
the angular power spectrum. 

\section{Conclusion}

	The aim of these lectures has been to understand the
physics responsible for the overall
properties of the standard model density and temperature
anisotropy power spectra (Figures 1 and 3).
Perturbations in the standard model are taken to be Gaussian,
and so these two quantities are the fundamental and unique
characterization of the perturbations.  An extremely
significant point is that the physics governing
perturbation evolution has imposed important scales on 
these power spectra.  These scales were not
present in the primordial spectrum, which we took
to be a pure power--law.  For the density perturbations, 
the imposed scale corresponds to 
the comoving horizon at the radiation--matter
transition (the equality epoch).   The final density power 
spectrum, thanks to this imposed scale, has roughly the correct 
shape to explain both the observed large--scale CMB fluctuations 
and the measured galaxy distribution today.  This is a
remarkable concordance over many orders of magnitude, a
true success for the basic theory.  In detail, the 
favored critical CDM model fails by a factor of two
over this vast range; other models with a  lower product of
$\Omega h$ are perfectly compatible with the present data.

	The scale imposed on the angular power spectrum
of the CMB anisotropies is the projected size of the
horizon at recombination.  The Doppler peaks
should first appear on this angular scale, and
because the projected angle 
depends on the curvature of space (e.g., Figure 4),
we have in principle a way to measure this fundamental
quantity.  Our discussion has focussed only on 
adiabatic perturbations, ignoring the possibility
of isocurvature modes.  The position of
the first Doppler peak is not exactly the same for these latter, 
but the basic principles
governing its location are.  In fact, the fine details of the Doppler
peaks depend on virtually all cosmological parameters,
leading to the hope that we may measure them 
with precision observations of the CMB; this has
become a preponderant subject of current cosmological
research. 
It is important to remark that this en-devour
will depend on the nature of the underlying model:
cosmic defect models certainly do not predict
the same dependence on cosmological parameters
as the standard model we have considered.
The CMB data will, of course, also tell us
which type of model is correct (at present,
it appears that defect models have difficulty
in explaining the CMB anisotropies and
the density power spectrum).  At any rate,
our brief discussion here is but an introduction
to an ever-growing subject which will take a 
center stage role in cosmology for the coming 
years.  \\

\vspace{1cm}

\noindent{\bf Acknowledgments}\\

	I would like thank the organizers and  
staff for a very enjoyable school, and the students
for their hard work.  Many thanks R. Durrer for discussions
of a factor of 2 over the course of two years! 
I also thank M. Douspis for his very nice anisotropy 
power spectrum plot.

General references are given in the Introduction.


\begin{thebibliography}{}
\bibitem{} Dicke R.H., Peebles P.J.E., Rolls P.G., Wilkinson D.T.
	1965, ApJ 142, 414
\bibitem{} Partridge R.B. 1995, 3K: The Cosmic Microwave
	Background Radiation, Cambridge Astrophysics Series,
	Cambridge University Press (Cambridge)
\bibitem{} Peebles P.J.E. \& Yu J.T. 1970, ApJ 162, 815
\bibitem{} Penzias A.A. \& Wilson R.W. 1965, ApJ 142, 419
\bibitem{} Sachs R.K. \& Wolfe A.M. 1967, ApJ 147, 73
\bibitem{} Smoot G.F. 1997, astro--ph/9705135
\bibitem{} Smoot G.F. et al. 1992, ApJ 396, L1
\end{thebibliography}
\end{document}